\RequirePackage{silence}
\PassOptionsToPackage{pdfpagelabels=false}{hyperref}
\documentclass[fleqn,usenatbib,useAMS]{mnras}
\usepackage{newtxtext,newtxmath}
\usepackage[T1]{fontenc}
\usepackage{ae,aecompl}

\usepackage{graphicx}	
\usepackage{amsmath}	

\DeclareRobustCommand{\appropto}{\mathrel{\vcenter{
			\offinterlineskip\halign{\hfil$##$\cr 
				\propto\cr\noalign{\kern2pt}\sim\cr\noalign{\kern-2pt}}}}}
\DeclareRobustCommand{\oversim}[2]{\protect{\mbox{\lower0.5ex\vbox{%
  \baselineskip=0pt\lineskip=0.2ex
  \ialign{$\mathsurround=0pt #1\hfil##\hfil$\crcr#2\crcr\sim\crcr}}}}}

\title[The Kennicutt-Schmidt law in MOND]{The Kennicutt-Schmidt law and the main sequence of galaxies in Newtonian and Milgromian dynamics}

\author[Zonoozi et al.]
{Akram Hasani Zonoozi$^{1,2}$\thanks{E-mail: \mbox{\href{mailto:a.hasani@iasbs.ac.ir}{a.hasani@iasbs.ac.ir}} }, Patrick Lieberz$^{2}$,
  Indranil Banik$^2$, Hosein Haghi$^{1}$ \newauthor and Pavel Kroupa$^{2,3}$ \\
$^{1}$Department of Physics, Institute for Advanced Studies in Basic Sciences (IASBS), PO Box 11365-9161, Zanjan, Iran\\
$^{2}$Helmholtz-Institut f\"ur Strahlen-und Kernphysik (HISKP), Universit\"at Bonn, Nussallee 14-16, D-53115 Bonn, Germany\\
$^{3}$Astronomical Institute, Faculty of Mathematics and Physics, Charles University in Prague, V Hole\v{s}ovi\v{c}k\'ach 2, CZ-180 00 Praha 8, Czech Republic\\}

\date{Accepted XXX. Received YYY; in original form ZZZ}

\pubyear{2021}

\begin{document}
\label{firstpage}
\pagerange{\pageref{firstpage}--\pageref{lastpage}}
\maketitle

\begin{abstract}

The Kennicutt-Schmidt law is an empirical relation between the star formation rate surface density ($\Sigma_{\text{SFR}}$) and the gas surface density ($\Sigma_{\text{gas}}$) in disc galaxies. The relation has a power-law form $\Sigma_{\text{SFR}} \propto  \Sigma_{\text{gas}}^{n}$. Assuming that star formation results from gravitational collapse of the interstellar medium, $\Sigma_{\text{SFR}}$ can be determined by dividing $\Sigma_{\text{gas}}$ by the local free-fall time $t_{\text{ff}}$. The formulation of $t_{\text{ff}}$ yields the relation between $\Sigma_{\text{SFR}}$ and $\Sigma_{\text{gas}}$, assuming that a constant fraction ($\varepsilon_{\text{SFE}}$) of gas is converted into stars every $t_{\text{ff}}$. This is done here for the first time using Milgromian dynamics (MOND). Using linear stability analysis of a uniformly rotating thin disc, it is possible to determine the size of a collapsing perturbation within it. This lets us evaluate the sizes and masses of clouds (and their $t_{\text{ff}}$) as a function of $\Sigma_{\text{gas}}$ and the rotation curve. We analytically derive the relation $\Sigma_{\text{SFR}} \propto  \Sigma_{\text{gas}}^{n}$ both in Newtonian and Milgromian dynamics, finding that $n=1.4$. The difference between the two cases is a change only to the constant pre-factor, resulting in increased $\Sigma_{\text{SFR}}$ of up to 25\% using MOND in the central regions of dwarf galaxies. Due to the enhanced role of disk self-gravity, star formation extends out to larger galactocentric radii than in Newtonian gravity, with the clouds being larger. In MOND, a nearly exact representation of the present-day main sequence of galaxies is obtained if $\epsilon_{\text{SFE}} = \text{constant} \approx 1.1\%$. We also show that empirically found correction terms to the Kennicutt-Schmidt law are included in the here presented relations. Furthermore, we determine that if star formation is possible, then the temperature only affects $\Sigma_{\text{SFR}}$ by at most a factor of $\sqrt{2}$.

\end{abstract}


\section{Introduction}
\label{Introduction}

Since stars form out of collapsing gas clouds, we expect a strong relation between the volume densities of star formation and gas. However, it is only possible to measure surface densities of external galaxies integrated along the line of sight, and even then the angular resolution is limited.  Therefore, the most commonly used relation \citep[known as the Kennicutt-Schmidt or KS law;][]{Schmidt, Kennicutt} relates the star formation rate (SFR) surface density $\Sigma_{\text{SFR}}$ to the gas surface density $\Sigma_{\text{gas}}$. The relation is assumed to be a power law of the form
\begin{equation}
	\Sigma_{\text{SFR}} \propto \Sigma_{\text{gas}}^{n} \, .
\end{equation}

The empirical relation between $\Sigma_{\text{SFR}}$ and $\Sigma_{\text{gas}}$ in a given region of a galaxy was first proposed by \citet{Schmidt}. This Schmidt law assumed $n \approx 2$ based on observations of the Solar neighbourhood. \citet{Kennicutt} used H$\alpha$ emission as a quantitative star formation tracer to determine the SFR both as a function of galactocentric distance and averaged over entire galactic discs. In the latter case, a strong correlation is obtained between averaged SFR and $\Sigma_{\text{gas}}$. Furthermore, \citet{Kennicutt} found that star formation appears to cease  where the Toomre stability criterion for a gas disc \citep{Toomre} indicates star formation to be impossible. In a second study, \citet{Kennicutt_1998} examined the correlation between disc-averaged SFRs and gas densities for a large sample of star forming galaxies, including 61 nearby, normal spiral galaxies and 36 starburst galaxies, finding a non-linear power-law function with index $n \approx 1.4 \pm 0.15$.

Since then, a value of $n\approx 1.5$ has become accepted as it is supported by later observations \citep[e.g.][]{Heyer, Leroy, Kennicutt2}, forming the basis for both theoretical and simulation work \citep[e.g.][]{Schaye}. These studies indicate a break in the $\Sigma_{\text{SFR}} \propto \Sigma_{\text{gas}}^{n}$ law such that $n > 1.5$ at small values of $\Sigma_{\text{gas}} \la 1 \, M_{\odot}/\text{pc}^2$ \citep[see also][]{Bigiel}, indicating the possible presence of a threshold gas density below which star formation becomes inefficient. In a galaxy, we can directly measure only the projected surface density, while the intrinsic volume density $-$ which is arguably more physically meaningful $-$ usually remains inaccessible. Recently, \citet{Bacchini_2019b, Bacchini_2019a} have developed and consolidated a method to convert the observed radial $\Sigma_{\text{gas}}(R_{\rm{gal}})$ and the radial $\Sigma_{\text{SFR}}(R_{\rm{gal}})$ profiles to the corresponding volume density profiles using the scale height of the galactic disc, i.e. its vertical distribution. They found that the volumetric star formation (VSF) law involving the SFR volume density and the total gas density $(HI+H_2)$ is a power-law relation with index $n\approx 2$, with no break, and a smaller intrinsic scatter than the $\Sigma_{\text{gas}}$-based star formation law. Moreover, \citet{Bacchini_2020} extended the VSF law to the regime of dwarf galaxies and the outskirts of spiral galaxies, which is of primary importance to investigate the presence of a density threshold for star formation in low-density, low-metallicity HI-dominated environments.

On the other hand, ultraviolet observations find a value of $n=0.99$ \citep{Boissier}. Compared to H$\alpha$ fluxes, ultraviolet observations are less sensitive to the presence of very massive young stars, and are therefore more likely to derive a value of $n$ close to the true one. Furthermore, \citet{PflammNature} support $n=1$ through the radially changing initial mass function of stars expected in the integrated galactic initial mass function (IGIMF) theory \citep{KroWei, Kroupabook}. In view of the discrepancy between the different exponents $n$, we readdress this problem theoretically within the basics of star formation in a galactic disc.

Since the free-fall timescale $t_\text{ff}$ depends on the gravity law, we study how changing the gravity law affects the value of $n$ in the KS law. Our main aim in this paper is to derive for the first time the KS law from a basic description in the framework of Milgromian dynamics \citep[MOND;][]{Milgrom_1983, FamaeyMcGaugh2012}. MOND is an alternative approach to a cold dark matter-dominated universe deduced from the flattening of observed rotation curves of spiral galaxies under the assumption of Newtonian dynamics. In MOND, these dynamical discrepancies are addressed by a generalization of Newtonian gravity \citep[for a thorough review, see e.g.][]{FamaeyMcGaugh2012}. Within the classical MOND framework, the Newtonian gravitational acceleration $g_{\rm{N}}$ is replaced in the spherically symmetric case by $g=\sqrt{g_{\rm{N}} a_0}$ when the gravitational acceleration is far smaller than the critical acceleration $a_0 = 1.2 \times 10^{-10}$ m/s$^2$ \citep{Begeman_1991, Gentile_2011}. In less symmetric configurations, the equations of motion are derived from a Lagrangian, yielding standard equations of motion but with a generalized Poisson equation for the gravitational field \citep{Lagrangian, Milgrom_2010}. MOND predicted the very tight radial acceleration relation (RAR) between the gravity $g$ implied by disc galaxy rotation curves and the Newtonian gravity $g_{\rm{N}}$ resulting from their baryonic distribution \citep{rar, Li_2018}. The RAR is also evident in stacked galaxy-galaxy weak lensing measurements that probe out to larger radii \citep{Milgrom_2013, Brouwer_2021}. The external field effect (EFE) predicted by MOND \citep{Lagrangian} and required for consistency with data on Solar neighbourhood wide binaries \citep{Banik_2018_Centauri, Pittordis_2019} has recently been confirmed at high significance by comparing galaxies in isolated and more crowded environments \citep{Haghi_2016, Chae_2020}. Detailed numerical simulations of disc galaxy secular evolution in MOND have recently been conducted for the case of M33 \citep{Banik_2020_M33} and for a Milky Way or M31-like surface density \citep{Roshan_2021}, while star formation has also been explored with high-resolution simulations \citep{Renaud_2016}. The possible cosmological context has been explored in e.g. \citet{Haslbauer_2020} and \citet{Asencio_2021}. The MOND corrections to Newtonian gravity might be capturing effects of the quantum vacuum \citep*{vacuum, vacuum2, Verlinde_2017, Senay_2021}.

We use the free-fall time as an approximation for the time the gas needs to collapse into stars. This principal idea has already been explored by e.g. \citet{Krumholz}. In contrast to their work, our approach is completely two-dimensional. We do not mix a two-dimensional surface density and a three dimensional free-fall time. This is achieved by assuming that the collapsing area of the disc can be calculated using a thin disc stability analysis \citep*{Toomre, MoNDdisc}.

Our paper is organized as follows: the basics of MOND and the EFE are discussed in Section \ref{MOND}. In Sections \ref{SFR_Newton} and \ref{SFR_MOND}, we describe the SFR surface density in Newtonian and Milgromian dynamics, respectively. Asymptotic limits are discussed in Section \ref{Extreme_SFR}, giving a good analytic understanding of our main results up to factors of order unity. Our main results are presented in Section \ref{data}. Finally, we discuss our results and conclude in Section \ref{Conclusions}.

\section{Basics of MOND}
\label{MOND}

Within the framework of MOND as derived from a classical Lagrangian \citep{Lagrangian}, the gravitational field strength $g$ from an isolated spherically symmetric mass distribution is \citep{Milgrom_1983}:
\begin{equation}
	g_{\rm N} ~=~ g \mu \left( \overbrace{\frac{g}{a_0}}^x \right) \, , \quad \text{or}~~
	g ~=~ g_{\rm{N}} \nu \left( \overbrace{\frac{g_{\rm{N}}}{a_0}}^y \right) \, ,
	\label{MOND_general}
\end{equation}
where $g$ and $g_{\rm N}$ are the MONDian and Newtonian gravitational accelerations due to the baryonic matter, respectively, and $a_0$ is the characteristic acceleration scale of MOND that is found to be $a_0 \approx 1.2 \times 10^{-10}\,{\rm m}/{\rm s}^2 \approx 3.8~\rm{kpc}\,\rm{Myr}^{-2}$. The interpolating function $\mu(x)$ or $\nu(y)$ has to reproduce Newtonian dynamics at large accelerations, i.e. $\mu(x) \rightarrow 1$ and $\nu(y) \rightarrow 1$ for $x \rightarrow \infty$ and $y \to \infty$. In the opposite limit of very small $x$ and $y$, the MOND behaviour is recovered if $\mu(x) \to x$ and $\nu(y) \rightarrow y^{-1/2}$. The relation between the $\mu$ and $\nu$ functions of MOND was discussed in detail in section 7.2 of \citet{Banik_2018_Centauri}.

MOND states that if the Newtonian gravitational acceleration $g_{\text{N}} \gg a_0$, the Newtonian gravitational description should be used. However, if $g_{\text{N}} \ll a_0$, the gravitational acceleration $g$ from an isolated point mass becomes
\begin{equation}
	g_{\rm M} ~=~ \sqrt{g_{\rm N} a_0} ~=~ \frac{\sqrt{G M a_0}}{R} \, .
	\label{MoND_simple}
\end{equation}
In other words, if the distance $R$ from an object of mass $M$ is much larger than its MOND radius $R_{\text{M}}$, then the gravitational acceleration will be given by Eq. \ref{MoND_simple} rather than by Newtonian gravity. $R_{\text{M}}$ follows from Eq. \ref{MoND_simple} for $g_{\rm M} = g_{\rm N} = a_0$:
\begin{equation}
	R_{\text{M}} ~=~ \sqrt{\frac{GM}{a_0}} \, .
	\label{R_M}
\end{equation}
In the case of a single point mass whose $g_{\text{N}} = GM/R^2$, Eq. \ref{MOND_general} would become
\begin{equation}
	g ~=~ \frac{G M}{R^2} \nu \left( \frac{G M}{a_0 R^2} \right) \, .
	\label{MOND_pointmass}
\end{equation}

There are three well-defined asymptotic cases in MOND:
\begin{enumerate}
	\item If $g_{\text{N}} \gg a_0$, we recover the classical Newtonian description because $\nu \to 1$ and so
	\begin{equation}
		g ~=~ g_{\text{N}} ~=~ \frac{G M}{R^2} \, .
		\label{Newtonian_approx}
	\end{equation}
	\item An isolated point mass with $g_{\text{N}} \ll a_0$ constitutes the isolated deep-MOND (idM) limit, where $\nu \left( y \right) \to 1/\sqrt{y}$. According to Eq. \ref{MOND_pointmass}, the gravitational acceleration becomes
	\begin{equation}
		g_{\text{idM}} ~=~ \sqrt{a_0 g_{\text{N}}} ~=~ \frac{\sqrt{G M a_0}}{R} \, .
	\end{equation}
	\item If the point mass is not isolated and the gravitational acceleration of an external field ($g_{\text{ext,N}}$) is much larger than the internal gravitational field ($g_{\text{int,N}}$) then we enter the quasi-Newtonian (QN) regime ($g_{\text{int,N}} \ll g_{\text{ext,N}}$). Averaged over the weak angular dependence, the result is \citep{Kfactor}:
	\begin{eqnarray}
	\label{MoND_ext}
	g_{\rm{QN}} ~&=&~ g_{\text{int,N}} \nu \left( \frac{g_{\text{ext,N}}}{a_0} \right) \left( 1 + \frac{K}{3} \right) \, , \\
	K ~&\equiv&~ \left. \frac{d \log  \nu \left( y \right) }{d \log y} \right|_{y = g_{\text{ext,N}}/a_0} \, .
	\end{eqnarray}
\end{enumerate}
In the extreme case of the deep-MOND quasi-Newtonian regime, $\nu = \sqrt{a_0/g_{\text{ext,N}}}$ and $K=-\frac{1}{2}$, so $\left( 1 + \frac{K}{3} \right) = \frac{5}{6}$ and the gravitational acceleration becomes
\begin{equation}
	g_{\text{QN}} ~=~ \frac{5}{6} \sqrt{\frac{a_0}{g_{\text{ext,N}}}} g_{\text{int,N}} ~=~ \frac{5}{6} \sqrt{\frac{a_0}{g_{\text{ext,N}}}} \frac{GM}{R^2} \, .
\end{equation}

\begin{figure}
	\begin{center}
		\includegraphics[width=100mm,height=80mm]{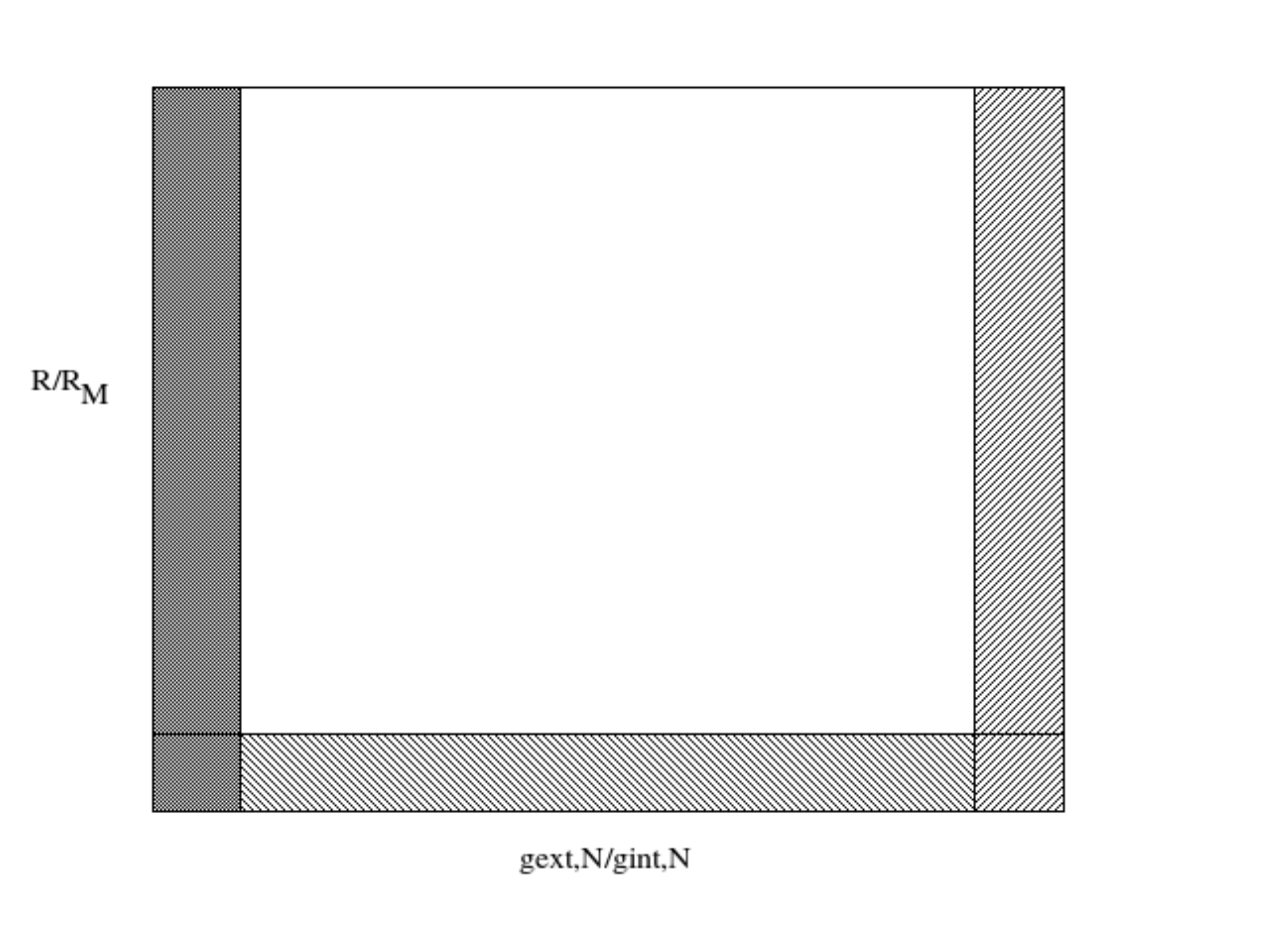}
		\caption{A schematic of the different MOND regimes. The $x$-axis represents the relative strength of the external gravitational acceleration $g_{\text{ext,N}}$ compared to the internal gravitational acceleration $g_{\text{int,N}}$, while the $y$-axis shows the distance in units of MOND radii $R_{\text{M}}$ (Eq. \ref{R_M}). The dark grey area on the left shows the isolated regime in which the interpolation function of Eq. \ref{MOND_pointmass} is valid for calculating the gravitational acceleration. The hatched part on the bottom of the schematic is completely in the Newtonian regime, so the gravitational acceleration can be calculated using Eq. \ref{Newtonian_approx}. Finally, the hatched area on the right is EFE-dominated and can therefore be calculated using Eq. \ref{MoND_ext}. There is no analytic solution for the upper middle (white) part of the schematic, where numerical methods are needed because both MOND and the EFE are important, but the EFE cannot be approximated as dominant}.
		\label{schematic}
	\end{center}
\end{figure}

To summarize: Eq. \ref{MOND_general} is valid for isolated objects, while Eq. \ref{MoND_ext} should be used if the external gravitational field is much stronger than the internal one. Note that if the combined Newtonian acceleration $g_{\text{N}}=\sqrt{g_{\text{int,N}}^2 + g_{\text{ext,N}}^2} \gg a_0$, then the system is in the Newtonian regime (for a summary, see Fig. \ref{schematic}). The EFE is unique to MOND due to its inherent non-linearity, which is required to fit the observed RAR.

\section{$\Sigma_{\text{SFR}}$ in Newtonian dynamics}
\label{SFR_Newton}

Assuming that star formation results from the radial collapse of the interstellar medium making up the galactic disc via short ($\approx 10$ Myr) lived giant molecular clouds (GMCs), we can express the SFR surface density as
\begin{equation}
	\Sigma_{\text{SFR}} ~=~ \frac{\Sigma_{\text{gas}}}{t_{\text{ff}}} \varepsilon_{\text{SFE}} \, ,
	\label{Ssfr}
\end{equation}
where $\Sigma_{\text{gas}}$ is the gas surface density at radius $R$ in the mid-plane of the galactic disc, $t_{\text{ff}}$ is the gravitational free-fall collapse time, and $\varepsilon_{\text{SFE}}$ is the star formation efficiency, i.e. the fraction of the gas cloud transformed into stars. It is well established that only a few percent of a GMC forms stars \citep[see e.g][]{zuckerman, protostars}. Moreover, GMCs constitute a small fraction of the ISM mass at any given time.

The Newtonian free-fall time $t_{\text{ff,N}}$ of a spherical cloud is \citep[see e.g.][]{Binney}:
\begin{equation}
	t_{\text{ff,N}} ~=~ \frac{\mathrm{\pi}}{2} \frac{R_{\text{cloud}}^{3/2}}{\sqrt{2 \, G M_{\text{cloud}}}} \, ,
	\label{tff}
\end{equation}
with $R_{\text{cloud}}$ being the radius and $M_{\text{cloud}}$ the mass of the cloud. Since $\Sigma_{\text{gas}}$ is known, $M_{\text{cloud}}$ can be expressed as
\begin{equation}
	M_{\text{cloud}} ~=~ \Sigma_{\text{gas}} \mathrm{\pi} R_{\text{cloud}}^{2} \, .
\end{equation}
Therefore,
\begin{equation}
	t_{\text{ff,N}} ~=~ \frac{\sqrt{\mathrm{\pi}}}{2} \sqrt{\frac{R_{\text{cloud}}}{2 \, G \Sigma_{\text{gas}}}} \, .
	\label{Free_fall_time_Newton}   
\end{equation}
Inserting this into Eq. \ref{Ssfr}, the Newtonian SFR surface density $\Sigma_{\text{SFR,N}}$ is:
\begin{equation}
	\Sigma_{\text{SFR,N}} ~=~ \frac{2}{\sqrt{\mathrm{\pi}}}\Sigma_{\text{gas}}^{3/2}\sqrt{\frac{2G}{R_{\text{cloud}}}} \varepsilon_{\text{SFE}} \, .
	\label{Ssfr2}
\end{equation}

The only remaining unknown is the cloud radius $R_{\text{cloud}}$. We determine $R_{\text{cloud}}$ using a stability criterion. Linear stability analysis of a uniformly rotating thin gas disc yields the following dispersion relation \citep{Binney, Escala}:
\begin{equation}
	\omega^{2} ~=~ \kappa^{2} - 2\mathrm{\pi} G \Sigma_{\text{gas}} \left| k \right| + C_{s}^{2} k^{2}  \, ,
	\label{omega}
\end{equation}
where $\omega$ is the oscillation frequency, $\kappa$ is the radial epicyclic frequency, $C_{s}$ is the isothermal sound speed, and $k \equiv 2\mathrm{\pi}/\lambda$ is the wavenumber, with $\lambda$ being the perturbation's wavelength. The region under study is stable against collapse if $\omega^{2} > 0$. For $\omega^{2}<0$, there is a growing mode, so perturbations grow exponentially and the disc is unstable. We thus set $\omega^{2}=0$ as the dividing line between stability/instability to find the critical wavelength $\lambda_{\rm{crit}}$. By approximating $R_{\text{cloud}} = \lambda_{\rm{crit}}/2$, we get that
\begin{eqnarray}
	R_{\text{cloud}} ~=~ \frac{\lambda_{\rm{crit}}}{2} ~=~ \frac{\mathrm{\pi}^{2} G \Sigma_{\text{gas}}}{\kappa^{2}} \pm \sqrt{\frac{\mathrm{\pi}^{4} G^{2} \Sigma_{\text{gas}}^{2}}{\kappa^{4}} - \frac{\mathrm{\pi}^{2}C_{s}^{2}}{\kappa^{2}}}  \, .
	\label{Rcloud}
\end{eqnarray}
Two different solutions $R_{\rm cloud,-}$ and $R_{\rm cloud,+}$ are possible for the radius of a region which is unstable against collapse.

\begin{figure*}
	\begin{center}
		\includegraphics[width=85mm,height=65mm]{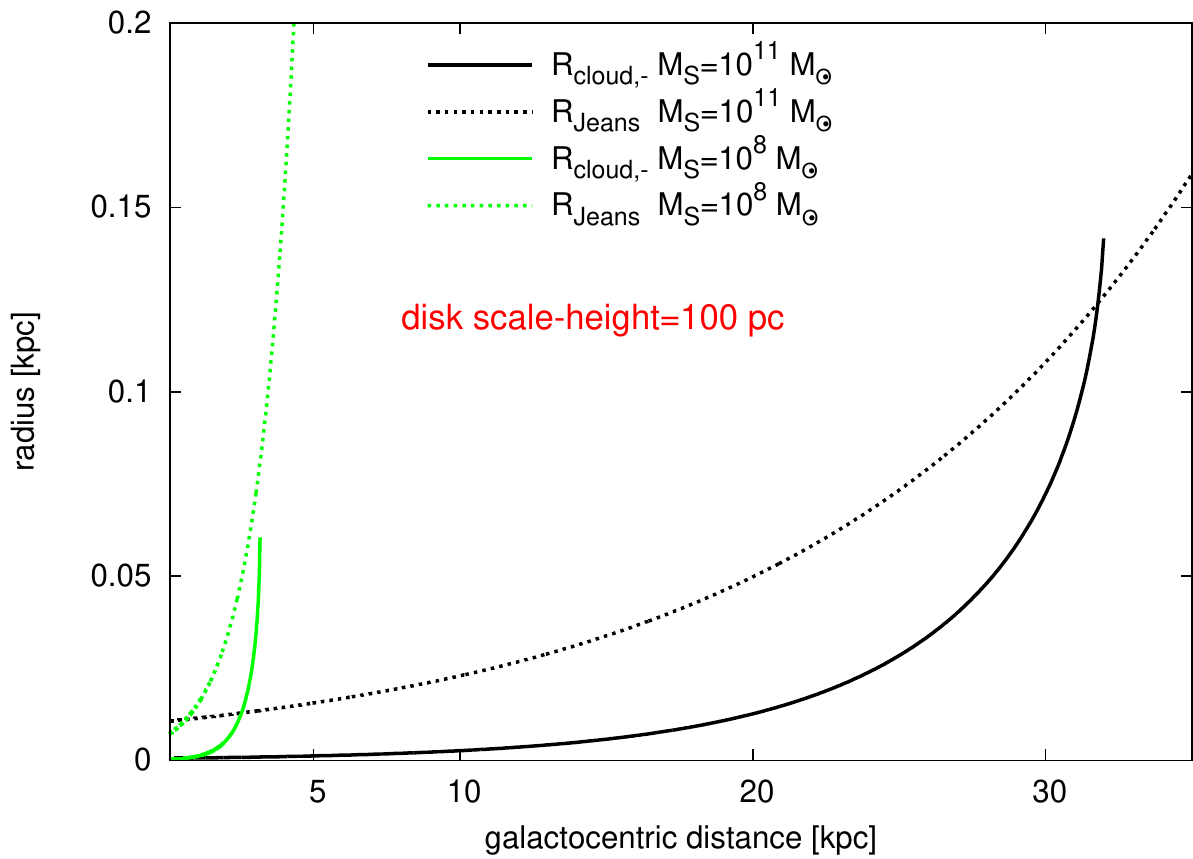}
		\includegraphics[width=85mm,height=65mm]{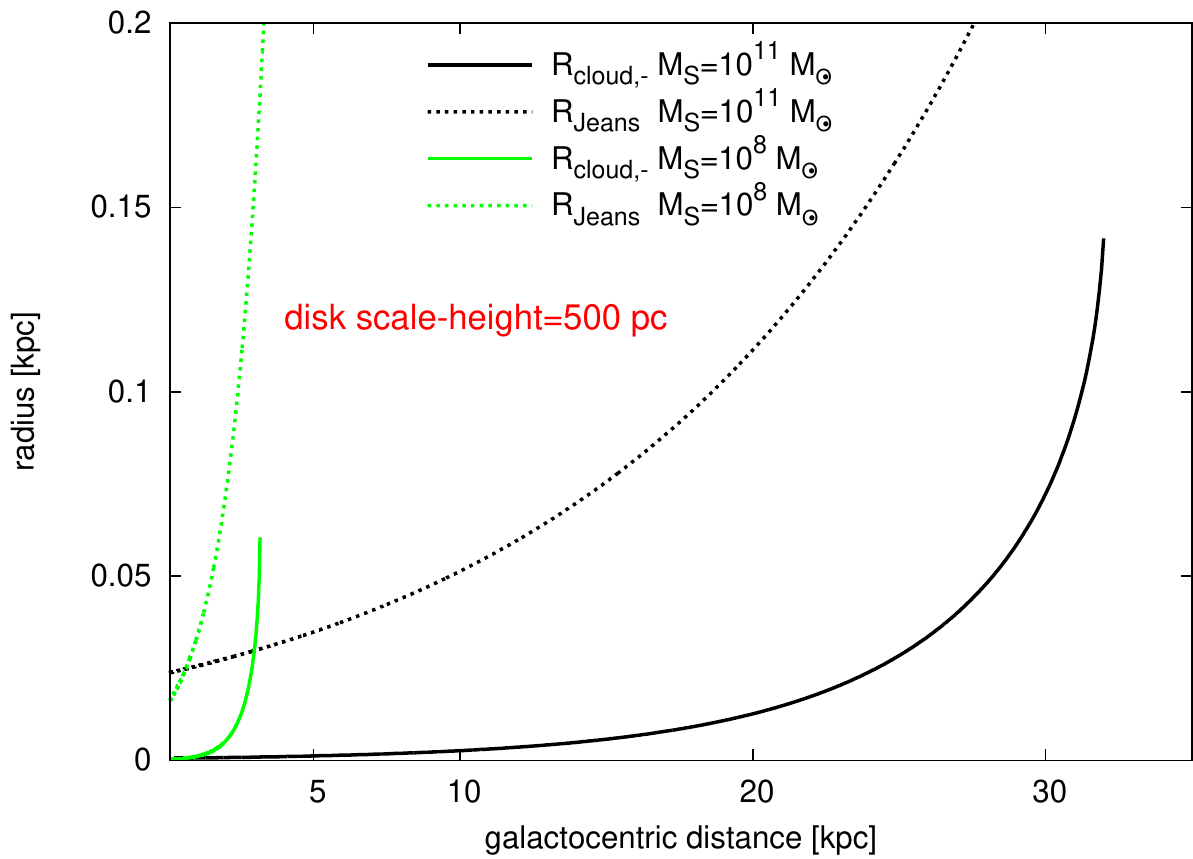}
		\includegraphics[width=85mm,height=65mm]{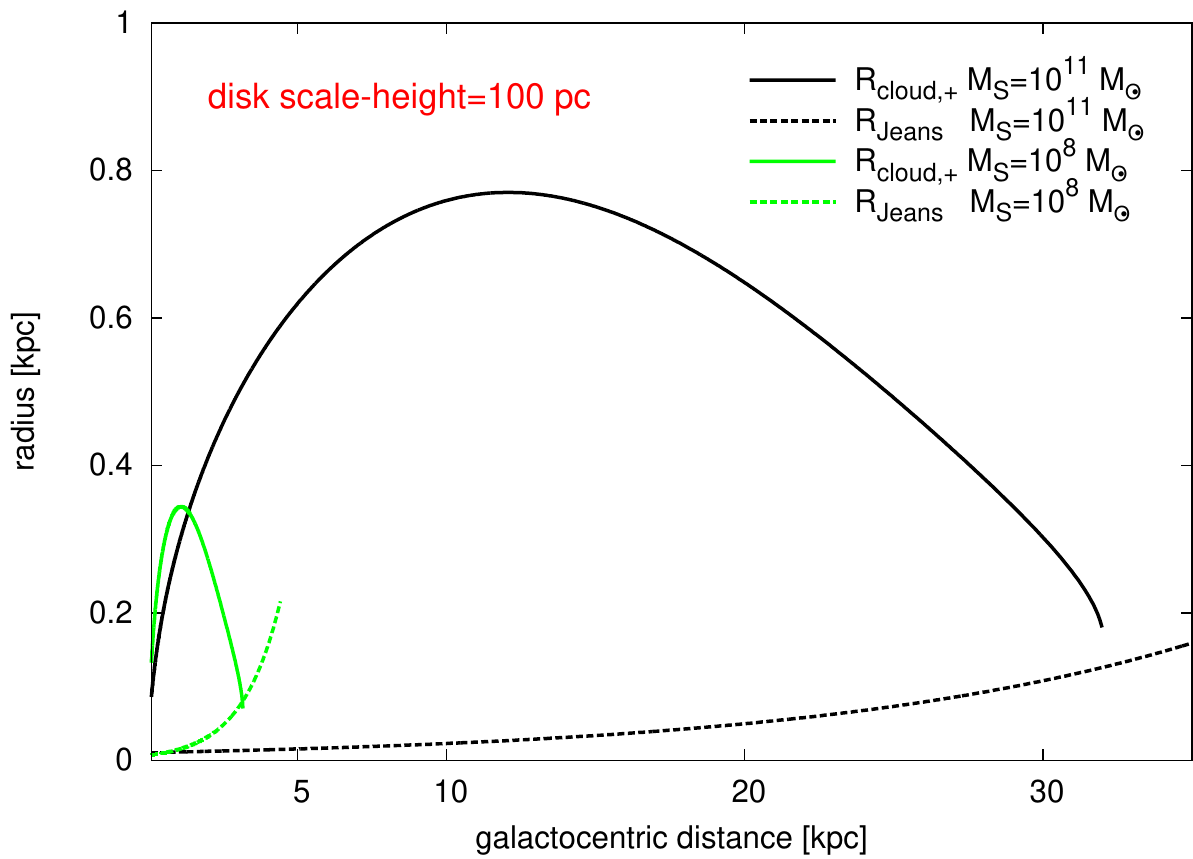}
		\includegraphics[width=85mm,height=65mm]{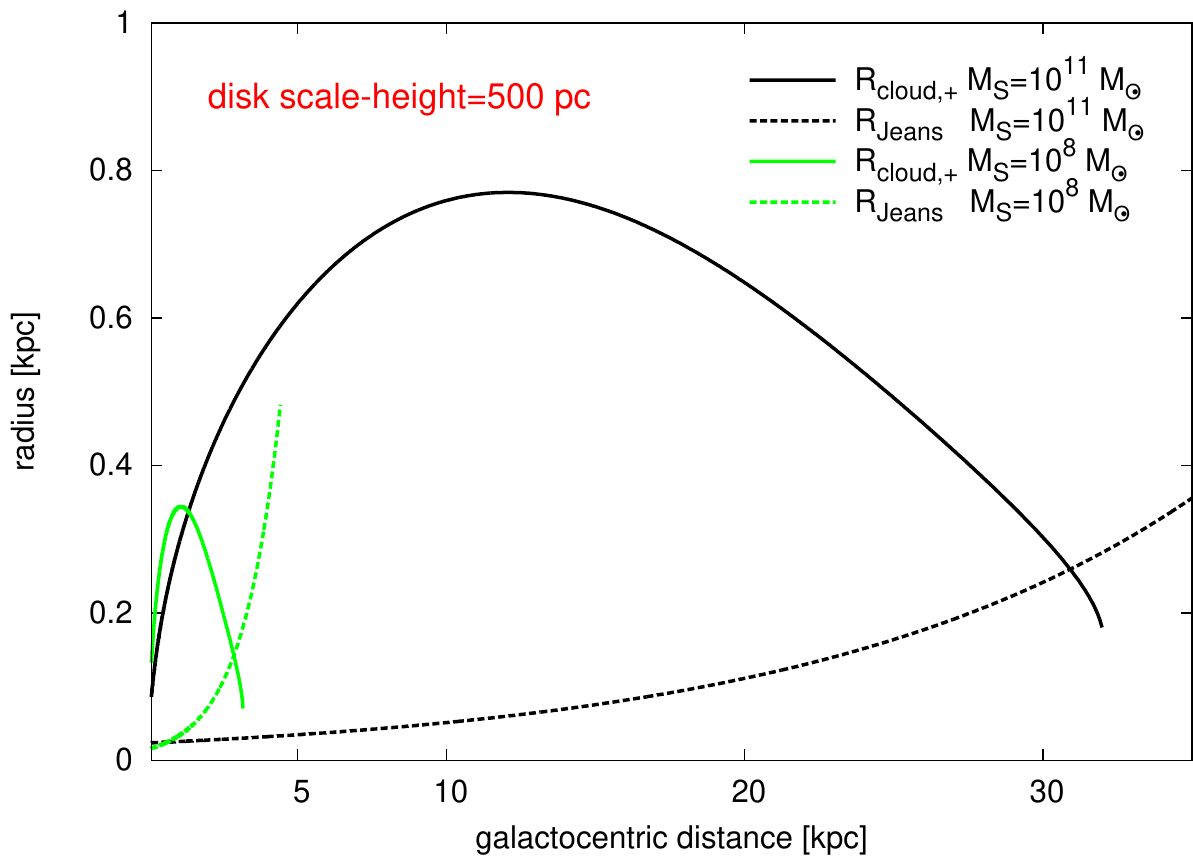}
		\caption{Upper panels: Cloud radius $R_{\rm cloud,-}$ (smaller solution to Eq. \ref{Rcloud}; solid lines) and Jeans radius (dashed lines) as a function of galactocentric radius in the Newtonian regime for two galaxies with $M_S = 10^8$ $M_{\odot}$ (green curves) and $M_S = 10^{11}$ $M_{\odot}$ (black curves). The left panels show results for a gas disk thickness of 100 pc, while the right panels assume 500 pc. The calculated cloud radius is unaffected by this assumption as it is based on a 2D analysis (see the text), but for a given surface density, changing the disk thickness changes the volume density and thus the Jeans length. Since $R_{\rm cloud,-}$ is smaller than the Jeans radius, these clouds are stable against collapse (compare curves with the same colour). Lower panels: same as upper panels, but for the higher solution to Eq. \ref{Rcloud} ($R_{\rm cloud,+}$). This solution is larger than the Jeans length.  We therefore focus on $R_{\rm cloud,+}$ elsewhere in this article. Notice also that the 2D collapse treatment used here is appropriate for a scale height of 100 pc, which we argue in the text is more realistic.}
		\label{Rcloud-RJeans}
	\end{center}
\end{figure*}

With the smaller solution $R_{\rm cloud,-}$, the cloud radius is much smaller than the disc thickness (Fig. \ref{Rcloud-RJeans}). In addition, the cloud can collapse only if its radius is larger than the Jeans radius
\begin{equation}
    \lambda_J ~=~ \frac{C_s}{\sqrt{G\rho}} \, ,
    \label{Jeans_wavelength}
\end{equation}
where $C_s$ is the speed of sound, which is in the range $0.2-0.7$ km/s for GMCs with an average temperature of $10-50$ K. We adopt $C_s = 0.5$ km/s when estimating the Jeans radius. This also requires an assumption about the vertical distribution.   In order to calculate the Jeans radius, according to Eq. \ref{Jeans_wavelength}, we use $\rho = \Sigma/h$ for two different values of the disk scale-height ( $h=100$ pc and $h=500$ pc). There are two possible solutions for the cloud radius. If we compare the resulting Jeans radius with $R_{\rm cloud,-}$ for two different galaxy stellar masses $M_S$, we see that throughout the galaxy, $R_{\rm cloud,-}$ is always smaller than the Jeans radius (Fig. \ref{Rcloud-RJeans}). Clouds are thus stable against collapse if we adopt $R_{\rm cloud,-}$.

However, $R_{\rm cloud,+}$ is larger than the Jeans radius for every galactocentric distance (lower panel of  Fig. \ref{Rcloud-RJeans}). Therefore, only the larger solution $R_{\rm cloud,+}$ corresponds to clouds that can collapse to form stars according to the Jeans criterion (Eq. \ref{Jeans_wavelength}). Note also that for a thin gas disc with a typical thickness of order 100 pc \citep{Nakanishi_2006, Kalberla_2008, Bacchini_2019b, Patra_2020}, the cloud radius is clearly larger. This shows that the 2D collapse treatment used here is appropriate. In what follows, we therefore focus on the solution $R_{\rm cloud,+}$.

Inserting $R_{\rm cloud,+}$ into Eq. \ref{Ssfr2}, the Newtonian $\Sigma_{\text{SFR,N}}$ becomes
\begin{equation}
	\Sigma_{\text{SFR,N}} ~=~ \varepsilon_{\text{SFE}} \frac{2}{\mathrm{\pi}^{3/2}} \kappa \Sigma_{\text{gas}}\sqrt{\frac{2}{1 + \sqrt{1-\frac{\kappa^{2} C_{s}^{2}}{\mathrm{\pi}^{2} G^2 \Sigma_{\text{gas}}^{2}}}}} \, .
	\label{Ssfrfinal}
\end{equation}
The radial epicyclic frequency $\kappa$ is \citep{Leroy2}:
\begin{equation}
	\kappa ~=~ \sqrt{2} \frac{V \left( R_{\text{gal}} \right)}{R_{\text{gal}}} \sqrt{1 + \frac{d \log{V \left( R_{\text{gal}} \right)}}{d \log{R_{\text{gal}}}}} \, ,
	\label{kappa}
\end{equation}
where $V \left( R_{\text{gal}} \right)$ is the rotation speed at radius $R_{\text{gal}}$. For a flat rotation curve, $\kappa \propto V \left( R_{\text{gal}} \right)/R_{\text{gal}} \propto 1/R_{\text{gal}}$, so $\Sigma_{\text{SFR}} \appropto \Sigma_{\rm gas}/R_{\rm gal}$ as long as star formation is possible.

The relation between $\Sigma_{\text{SFR}}$ and $\Sigma_{\text{gas}}$ is in agreement with the empirically found additional factors to the KS law. \citet{Prantzos} found a radial dependence to the KS law of $1/R_{\text{gal}}$ for both $n = 1$ and $n = 2$ (see their eq. 6, variable names changed for consistency):
\begin{equation}
	\Sigma_{\text{SFR}} ~=~ 0.3 \, \Sigma_{\text{gas}} (R_{\text{gal}}/R_{\odot})^{-1} \text{ M}_{\odot} \text{ pc}^{-2} \text{ Gyr}^{-1} \, .
\end{equation}
Based on a dynamical model, \citet{Boissier} also described an additional dependence on the galactic rotation curve of $V \left( R_{\text{gal}} \right)/R_{\text{gal}}$ \citep[eq. 6 from][variable names changed for consistency]{Boissier}:
\begin{equation}
	\Sigma_{\text{SFR}} ~=~ \alpha \Sigma^n_{\text{gas}} \frac{V \left( R_{\text{gal}} \right)}{R_{\text{gal}}}\, ,
\end{equation}
where $\alpha$ is a constant. Thus, the additional factor of \citet{Boissier} is equal to the one found by \citet{Prantzos} when the rotation curve becomes flat, and simply reflects that the free-fall time is of order $1/\kappa$.

\section{$\Sigma_{\text{SFR}}$ in MOND}
\label{SFR_MOND}

\begin{figure}
	\begin{center}
	\includegraphics[width=85mm,height=75mm]{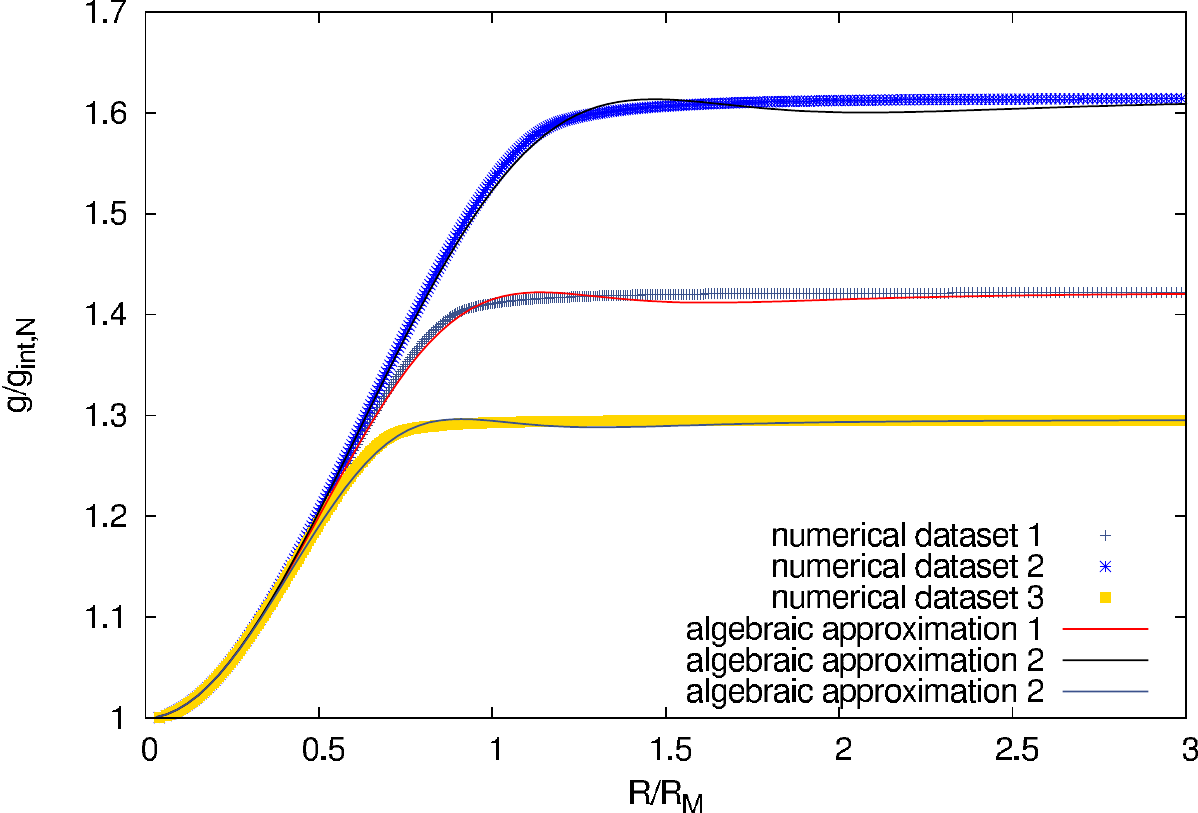}
	\caption{The numerical datasets from \citet{Kfactor} compared to the algebraic approximation (Eq. \ref{alg_approx}). The $x$-axis is the separation in MOND radii $R_{\rm M}$ (Eq. \ref{R_M}), while the $y$-axis is the angle-averaged radial gravitational acceleration calculated using MOND and shown relative to the Newtonian gravitational acceleration. The numerical dataset and algebraic approximation 1 refer to an external field roughly as strong as the Solar System is experiencing from the Milky Way \citep[see also][]{Klioner_2021}. Dataset and approximation 2 (3) are for an external field 0.7 (1.4) times stronger than in dataset 1.}
	\label{alg_approx_plot}
	\end{center}
\end{figure}

Since MOND is a non-Newtonian classical theory of gravitation, the free-fall time differs from that in Newtonian dynamics (Eq. \ref{Free_fall_time_Newton}). As pointed out in Section \ref{MOND}, the algebraic approximation of the MOND force law needs to be corrected in the case of a dominating external gravitational field by a factor of $\left( 1 + \frac{K}{3} \right)$. Thus, the first step in calculating a general Milgromian free-fall time is to derive a general algebraic description of the MOND force law. For this, we use the data from \citet{Kfactor}. A general correction factor of $\left(1 + \tanh\left(0.825 \frac{g_{\text{ext,N}}}{g_{\text{int,N}}} \right)^{3.7} \frac{K}{3}\right)$ manages to reproduce their numerical results sufficiently well (Fig. \ref{alg_approx_plot}). This factor becomes 1 in the Newtonian regime or if the external field is much weaker than the internal field, but is $\left( 1 + \frac{K}{3} \right)$ if the external field is much stronger than the internal field. It is therefore able to reproduce every asymptotic regime (grey regions) in the schematic of Fig. \ref{schematic}, but can be applied outside these limiting circumstances. Thus, a general algebraic approximation to the MOND force law takes the form
\begin{eqnarray}
	g ~=~ g_{\text{int,N}} \nu\left( \frac{g_{\text{N}}}{a_0} \right) \left(1 + \tanh\left(0.825 \frac{g_{\text{ext,N}}}{g_{\text{int,N}}} \right)^{3.7} \frac{K}{3} \right) \, ,
	\label{alg_approx}
\end{eqnarray}
where $g_{\rm {int,N}} = GM_{\rm cloud}/R_{\rm cloud}^2$ is the internal gravitational field, and the Newtonian gravitational field entering the calculation of $\nu$ and $K$ is
\begin{eqnarray}
	g_{\text{N}} ~=~ \sqrt{g_{\text{int,N}}^2 + g_{\text{ext,N}}^2} \, .
	\label{gN_def}
\end{eqnarray}

It is not possible to write down a closed algebraic description of the general free-fall time $t_{\text{ff}}$ in MOND valid for any state between the Newtonian regime, isolated MOND regime, and quasi-Newtonian regime. However, an approximate algebraic description is possible (see Appendix \ref{tff_derive} for the derivation):
\begin{equation}
	t_{\text{ff}} ~=~ \frac{\mathrm{\pi}}{2}\frac{R^{3/2}}{\sqrt{2GM\nu \left( g_{\text{N}}/a_0 \right) \left(1+\tanh\left(0.825 \frac{g_{\text{ext,N}}}{g_{\text{int,N}}} \right)^{3.7} \frac{K}{3}\right)}} N \, ,
	\label{tffMOND}
\end{equation} 
with 
$N$ being a numerical correction that becomes $\frac{2}{\sqrt{\mathrm{\pi}}}$ in the idM case and $1$ otherwise.
\begin{equation}
	N ~=~ 1 + \frac{\left( 1 - \frac{1}{\nu \left( g_{\text{N}}/a_0 \right)} \right) \left( \frac{2}{\sqrt{\mathrm{\pi}}} - 1 \right)}{\frac{g_{\text{ext,N}}}{g_{\text{int,N}}}+1}  \, .
	\label{N}
\end{equation} 

Note that in the Newtonian regime ($g_{\text{N}} \gg a_0 \Longrightarrow \nu\left( g_{\text{N}}/a_0 \right) \to 1 \, , ~K \to 0$), $t_{\text{ff}}$ becomes equal to $t_{\text{ff,N}}$ (Eq. \ref{tff}).

In the idM regime ($ g_{\text{ext,N}} \ll g_{\text{int,N}} \ll a_0 \Longrightarrow \nu \left( g_{\text{N}}/a_0 \right) \to \sqrt{\frac{a_0}{g_{\text{N}}}} \, , ~ \tanh\left(0.825\frac{g_{\text{ext,N}}}{g_{\text{int,N}}} \right)^{3.7} \to 0$), $t_{\text{ff}}$ simplifies to
\begin{equation}
	t_{\text{ff,idM}} ~=~ \sqrt{\frac{\mathrm{\pi}}{2}}\frac{R_{\text{cloud}}}{\sqrt[4]{G M_{\text{cloud}} a_0}} ~=~ \sqrt[4]{\frac{\mathrm{\pi}}{G \Sigma_{\text{gas}} a_0}} \sqrt{\frac{R_{\text{cloud}}}{2}} \, .
	\label{tffdeepMOND}
\end{equation}
This matches eq. 24 from \citet{MoNDfft}, as expected.

\begin{figure}
	\begin{center}
	\includegraphics[width=85mm,height=75mm]{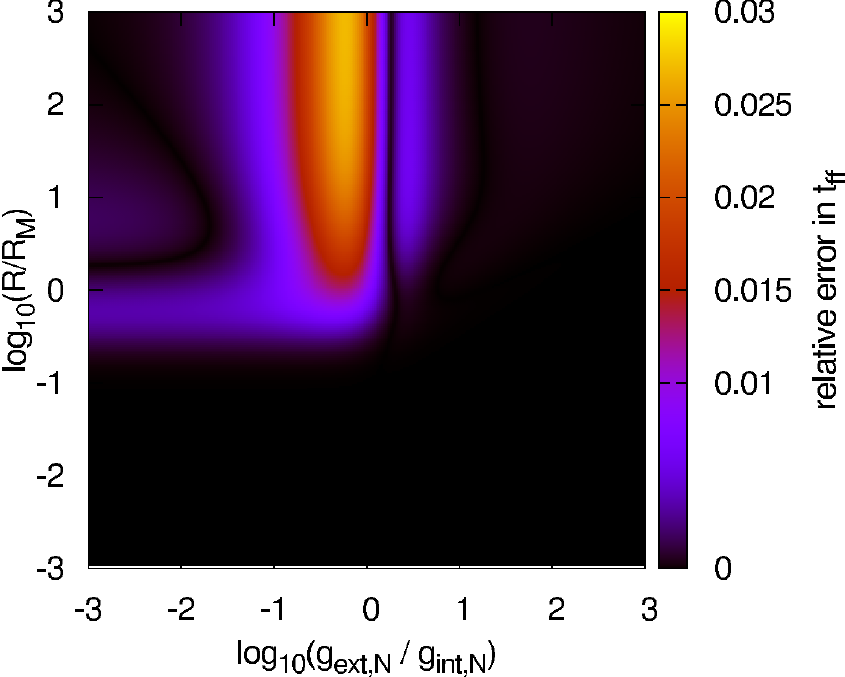}
	\caption{The relative error of our approximation (Eq. \ref{tffMOND}) for the free-fall time. The $x$-axis is the initial ratio of the external and internal gravitational fields, while the $y$-axis is the starting distance in MOND radii (Eq. \ref{R_M}). The maximum error of this approximation is only about 3\%.}
	\label{fft_approx}
	\end{center}
\end{figure}

In the quasi-Newtonian regime ($ g_{\text{int,N}} \ll g_{\text{ext,N}} \ll  a_0 \Longrightarrow \nu\left( g_{\text{N}}/a_0 \right) \to \sqrt{\frac{a_0}{g_{\text{ext,N}}}} \, ,~ \tanh\left(0.825\frac{g_{\text{ext,N}}}{g_{\text{int,N}}} \right)^{3.7} K \to -1/2$), $t_{\text{ff}}$ simplifies to
\begin{equation}\label{tffQN}
\begin{aligned}
 & t_{\text{ff,QN}} ~=~ \sqrt{\frac{3}{5}}\frac{\mathrm{\pi}}{2} \frac{R_{\text{cloud}}^{3/2}}{\sqrt[4]{\frac{a_0}{g_{\text{ext,N}}}} \sqrt{G M_{\text{cloud}}}} = \\
 & \frac{1}{2} \sqrt{\frac{3 \mathrm{\pi}}{5}} \sqrt{\frac{R_{\text{cloud}}}{\sqrt{\frac{a_0}{g_{\text{ext,N}}}} G \Sigma_{\text{gas}}} } = \frac{1}{2} \sqrt{\frac{3 \mathrm{\pi}}{5}} \sqrt{\frac{R_{\text{cloud}} V_\text{N}}{\sqrt{a_0 R_{\text{gal}}} G \Sigma_{\text{gas}}} } \, ,
\end{aligned}
\end{equation}
assuming $g_{\text{ext,N}} = V_\text{N}^2/R_{\text{gal}}$ is the Newtonian external field in the last part of the equation, $V_\text{N}$ is the Newtonian rotational velocity of the disc, and $R_{\text{gal}}$ is the distance to the galactic centre. For intermediate states, there is a divergence with the numerically determined $t_{\text{ff}}$ of only up to 3\% (Fig. \ref{fft_approx}).

In MOND, the dispersion relation of a thin gas disc changes to \citep{MoNDdisc}:
\begin{equation}
	\omega^{2} ~=~ \kappa^{2} - 2\mathrm{\pi} G \nu \left( g_{\text{N,disc}}/a_0 \right) \left(1 + \frac{K}{2}\right) \Sigma_{\text{gas}} \left| k \right| + C_{s}^{2} k^{2}  \, ,
	\label{omegaMOND}
\end{equation}
with $g_{\text{N,disc}}$ being the disc's Newtonian gravitational acceleration just above the disc \citep[see e.g.][]{Brada, MoNDdisc}:
\begin{equation}
	g_{\text{N,disc}} ~=~ \sqrt{g_{\text{N,r}}^2 + g_{\text{N,z}}^2} ~=~ \sqrt{\left( V_\text{N}^2/R_{\text{gal}} \right)^2 + \left( 2 \mathrm{\pi} G \Sigma_{\text{gas}} \right)^2} \, ,
	\label{gNdisc}
\end{equation}
where $g_{\text{N,r}}$ is the radial gravitational acceleration and $g_{\text{N,z}}$ the vertical one orthogonal to the galactic disc, with only the gas surface density entering into $g_{\text{N,z}}$ because the gas disk is typically much thinner than the stellar disk. We use N subscripts to denote Newtonian values. Note that in Eq. \ref{omegaMOND}, $K$ is without the $\tanh\left(0.825\frac{g_{\text{ext,N}}}{g_{\text{int,N}}} \right)^{3.7}$ pre-factor because the initial perturbation to the disc is always external field dominated \citep[see][]{MoNDdisc}.

The above equations result in a radial extent to the collapsing region of
\begin{eqnarray}
	R_{\text{cloud}} ~&=&~  \frac{\mathrm{\pi}^{2} G \nu \left( g_{\text{N,disc}}/a_0 \right) \left(1+\frac{K}{2}\right) \Sigma_{\text{gas}}}{\kappa^{2}} \\
	&+& \sqrt{\frac{\mathrm{\pi}^{4} G^{2} \nu^{2} \left( g_{\text{N,disc}}/a_0 \right) \left(1 + \frac{K}{2}\right)^{2} \Sigma_{\text{gas}}^{2}}{\kappa^{4}}-\frac{\mathrm{\pi}^{2}C_{s}^{2}}{\kappa^{2}}} \, . \nonumber
\end{eqnarray}
In general, the MOND star formation rate surface density therefore becomes
\begin{equation}
\label{SSFRfinal_gen}
\begin{aligned}
    &	\Sigma_{\text{SFR,M}} ~=~ \varepsilon_{\text{SFE}} \frac{2}{N \mathrm{\pi}^{3/2}}\kappa \Sigma_{\text{gas}} \times\\
    	& \sqrt{\frac{2 \nu \left( g_{\text{N}}/a_0 \right) \left(1 + \tanh\left(0.825 \frac{g_{\text{ext,N}}}{g_{\text{int,N}}} \right)^{3.7} \frac{K}{3}\right)}{\nu \left( g_{\text{N,disc}}/a_0 \right) \left(1 + \frac{K}{2}\right) + \sqrt{ \nu^{2} \left( g_{\text{N,disc}}/a_0 \right) \left(1+\frac{K}{2}\right)^{2}-\frac{\kappa^{2} C_{s}^{2}}{\mathrm{\pi}^{2} G^2 \Sigma_{\text{gas}}^{2}}}}} \, .
\end{aligned}	
\end{equation}
In the Newtonian limit, this is the same as Eq. \ref{Ssfrfinal}. In the isolated-deep MOND limit (which arises near the centre as it requires $g_{\text{N,r}} \ll g_{\text{N,z}}$ to avoid external field domination), we get:
\begin{equation}
	\Sigma_{\text{SFR,idM}}=\varepsilon_{\text{SFE}}\frac{2}{\mathrm{\pi}} \frac{\sqrt[4]{2}}{\sqrt{3}} \kappa \Sigma_{\text{gas}}  \sqrt{\frac{2}{1 + \sqrt{1-\frac{32 \kappa^{2} C_{s}^{2}}{9 \mathrm{\pi} G \Sigma_{\text{gas}}}}}} \, .
\end{equation}
In the external field dominated deep-MOND case (which arises further out as it requires $g_{\text{N,z}} \ll g_{\text{N,r}} \ll a_0$), we get:
\begin{equation}
	\Sigma_{\text{SFR,QN}} = \varepsilon_{\text{SFE}}\frac{2\sqrt{10}}{3 \mathrm{\pi}^{3/2}} \kappa \Sigma_{\text{gas}} \sqrt{\frac{2}{1 + \sqrt{1 - \frac{16 V^2  \kappa^{2} C_{s}^{2}}{9 a_0 R_{\text{gal}} \mathrm{\pi}^{2} G^{2} \Sigma_{\text{gas}}^{2}}}}}\, .
\end{equation}

\section{Extreme cases of $\Sigma_{\text{SFR}}$}
\label{Extreme_SFR}

There are now two extreme solutions to Eq. \ref{SSFRfinal_gen} depending on the temperature $T$.

\subsection{Zero temperature disc}
\label{Zero_temperature_disc}

Mathematically, the lowest possible $C_{s} = 0$. As $C_{s}^{2} \propto T$, this would require an unphysical temperature of 0 K. Nonetheless, assuming $C_{s} = 0$ shows one of the extreme solutions to Eq. \ref{Ssfrfinal}. In this case, Eq. \ref{SSFRfinal_gen} simplifies to:
\begin{equation} \label{MoND_Cs0}
	\begin{aligned}
	\Sigma_{\text{SFR,M}} &= \varepsilon_{\text{SFE}} \frac{2}{N \mathrm{\pi}^{3/2}} \kappa \Sigma_{\text{gas}} \times \\
	&\sqrt{\frac{\nu \left( g_{\text{N}}/a_0 \right) \left(1+\tanh\left(0.825 \frac{g_{\text{ext,N}}}{g_{\text{int,N}}} \right)^{3.7} \frac{K}{3}\right)}{\nu \left( g_{\text{N,disc}}/a_0 \right) \left(1 + \frac{K}{2} \right)}} \, ,
	\end{aligned}
\end{equation}
which in the different cases becomes:
\begin{eqnarray}
	\label{Newtonextreme1}
	\Sigma_{\text{SFR,N}} &=& \varepsilon_{\text{SFE}}\frac{2}{\mathrm{\pi}^{3/2}} \kappa \Sigma_{\text{gas}} ~ \text{(Newtonian),} \\
	\Sigma_{\text{SFR,idM}} &=& \varepsilon_{\text{SFE}}\frac{2}{\mathrm{\pi}} \frac{\sqrt[4]{2}}{\sqrt{3}} \kappa \Sigma_{\text{gas}} ~ \text{(isolated deep-MOND),} \\
	\Sigma_{\text{SFR,QN}} &=& \varepsilon_{\text{SFE}}\frac{2\sqrt{10}}{3 \mathrm{\pi}^{3/2}}  \kappa \Sigma_{\text{gas}} ~ \text{(EFE dominated).}
	\label{extextreme1}
\end{eqnarray}
The numerical factors are 0.359, 0.437, and 0.378 for the Newtonian, idM, and QN regimes, respectively.

\subsection{Maximum temperature star-forming disc}
\label{Maximum_temperature_disc}

If $\kappa C_{s} > \mathrm{\pi} G \nu \left( g_{\text{N,disc}}/a_0 \right) \left(1+\frac{K}{2}\right) \Sigma_{\text{gas}}$, then Eq. \ref{SSFRfinal_gen} becomes imaginary, indicating that star formation is impossible ($\Sigma_{\text{SFR,M}}$ drops to 0). This happens at high temperature as $C_{s}^{2} \propto T$. If $C_{s}$ has the highest value consistent with star formation $\left( \kappa C_s = \mathrm{\pi} G \nu \left( g_{\text{N,disc}}/a_0 \right) \left(1+\frac{K}{2}\right) \Sigma_{\text{gas}} \right)$, then the other extreme case is:
\begin{equation} \label{MoND_Csn0}
	\begin{aligned}
	\Sigma_{\text{SFR,M}} ~&=&~ \varepsilon_{\text{SFE}}\frac{2 \sqrt{2}}{N \mathrm{\pi}^{3/2}} \kappa \Sigma_{\text{gas}} \\
	&\times& \sqrt{\frac{\nu \left( g_{\text{N}}/a_0 \right) \left(1+\tanh\left(0.825 \frac{g_{\text{ext,N}}}{g_{\text{int,N}}} \right)^{3.7} \frac{K}{3}\right)}{\nu \left( g_{\text{N,disc}}/a_0 \right) \left(1 + \frac{K}{2}\right)}} \, ,
	\end{aligned}
\end{equation}
which becomes
\begin{eqnarray} \label{Newtonextreme2}
	\Sigma_{\text{SFR,N}} &=& \varepsilon_{\text{SFE}}\frac{2 \sqrt{2} }{\mathrm{\pi}^{3/2}} \kappa \Sigma_{\text{gas}} ~ \text{(Newtonian)}, \\
	\Sigma_{\text{SFR,idM}} &=& \varepsilon_{\text{SFE}}\frac{2^{7/4}}{\mathrm{\pi}\sqrt{3}} \kappa \Sigma_{\text{gas}}  ~ \text{(isolated deep-MOND)}, \\
	\Sigma_{\text{SFR,QN}} &=& \varepsilon_{\text{SFE}}\frac{2\sqrt{20}}{3 \mathrm{\pi}^{3/2}}  \kappa \Sigma_{\text{gas}}  ~ \text{(EFE dominated)}.
	\label{extextreme2}
\end{eqnarray}
Notice that the numerical factors here exceed those in Section \ref{Zero_temperature_disc} by a factor of $\sqrt{2}$ in all three analytically tractable regimes.

\subsection{Intermediate temperature discs}

In the following, except if otherwise stated, we assume that star formation is possible. Therefore, $\Sigma_{\text{SFR}}$ must lie between the two extreme solutions to Eq. \ref{SSFRfinal_gen}, which only differ by a factor of $\sqrt{2}$ (compare Eqs. \ref{MoND_Cs0} and \ref{MoND_Csn0}). Eq. \ref{SSFRfinal_gen} can therefore be simplified to:
\begin{eqnarray}
	\label{finalwithW}
	\Sigma_{\text{SFR,M}} &=& \varepsilon_{\text{SFE}}\frac{2}{N \mathrm{\pi}^{3/2}} \kappa \Sigma_{\text{gas}} W \\
	&\times& \sqrt{\frac{\nu \left( g_{\text{N}}/a_0 \right) \left(1+\tanh\left(0.825 \frac{g_{\text{ext,N}}}{g_{\text{int,N}}} \right)^{3.7} \frac{K}{3}\right)}{\nu \left( g_{\text{N,disc}}/a_0 \right) \left(1 + \frac{K}{2}\right)}}  \, , \nonumber
\end{eqnarray}
where the factor
\begin{equation}
	W ~=~ \sqrt{\frac{2}{1 + \sqrt{1 - \frac{C_{s}^{2} \kappa^{2}}{\mathrm{\pi}^{2} G^{2}  \nu \left( g_{\text{N,disc}}/a_0 \right) \left(1 + \frac{K}{2}\right) \Sigma_{\text{gas}}^{2}}}}} ~=~ \left[ 1, \sqrt{2} \right]\, .
\end{equation}
Therefore, the two extremes only differ by a factor of $\sqrt{2}$ (factor $W$ from Eq. \ref{finalwithW}). Furthermore, $W$ is the only part of Eq. \ref{finalwithW} depending on the sound speed $C_{s}$, which is the only part of the equation depending on the temperature. Therefore, as long as star formation is possible (Eq. \ref{SSFRfinal_gen} has real solutions), the temperature only influences the star formation rate by at most a factor of $\sqrt{2}$.

\begin{figure}
	\begin{center}
	\includegraphics[width=87mm,height=75mm]{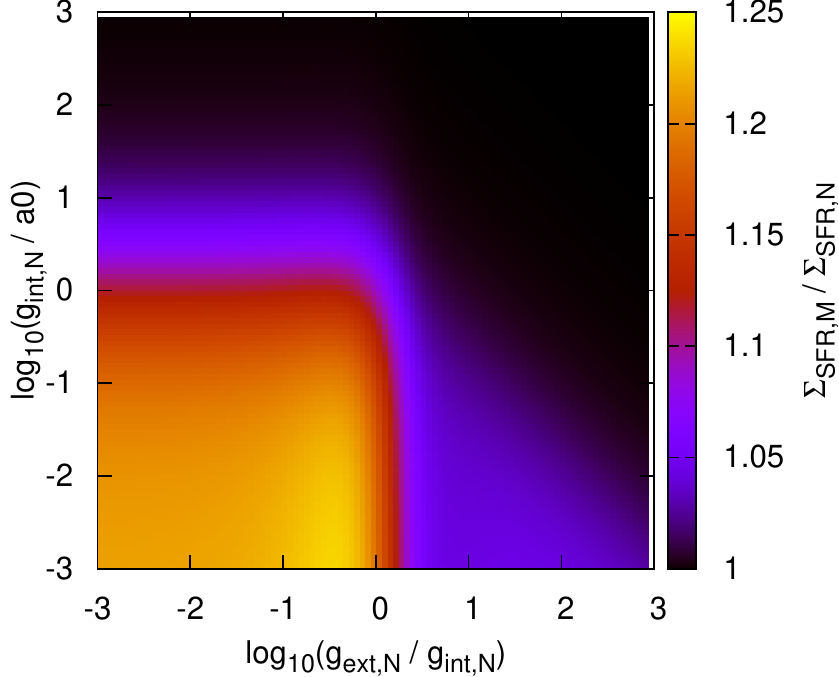}
	\caption{Ratio of the MOND $\Sigma_{\text{SFR}}$ (Eq. \ref{SSFRfinal_gen}) to the Newtonian result (Eq. \ref{Ssfrfinal}). The $x$-axis is the initial ratio of the external and internal Newtonian gravitational fields, while the $y$-axis is the initial Newtonian gravitational acceleration from the internal field in units of $a_0$. The maximal deviation of $\approx 25\%$ arises in the idM limit (towards the lower left).}
	\label{ssfr_dev}
	\end{center}
\end{figure}

As is evident from the extreme solutions (see Eqs. \ref{Newtonextreme1}-\ref{extextreme1} and Eqs. \ref{Newtonextreme2}-\ref{extextreme2}), the general dependence of $\Sigma_{\text{SFR}}$ on $\Sigma_{\text{gas}}$ does not change when going from the Newtonian to the idM regime or the external field dominated regime. Only the numerical pre-factors vary slightly. Eqs. \ref{MoND_Cs0} and \ref{MoND_Csn0} show why $-$ the MONDian factor $\nu$ contributes to both the free-fall time and the disc stability, leading to it cancelling out. Since $2 g_{\text{int,N}} = g_{\text{N,z}}$ and $g_{\text{ext,N}} = g_{\text{N,r}}$, only constant numerical contributions of order unity remain from the division of the $\nu$ parameters (Eq. \ref{SSFRfinal_gen}). The numerical deviation between the Newtonian and Milgromian formulations is up to 25\% (Fig. \ref{ssfr_dev}). A more significant difference arises in the sizes of collapsing regions, because $\Sigma_{\rm{gas}}$ in Eq. \ref{Rcloud} is enhanced in MOND by a factor of $\nu \left( 1 + \frac{K}{2} \right)$, which can be very large. This may lead to more massive star clusters further out in the disk.

\section{Comparison with the observed main sequence of star-forming galaxies}
\label{data}

\subsection{Calculating the rotation speed from SPARC data}
\label{SPARC_scaling_relations}

\citet*{SPARC} published the `\textit{Spitzer} Photometry and Accurate Rotation Curves (SPARC) galaxy sample, which consists of 175 galaxies with surface photometry at 3.6 $\mu\text{m}$ and extended HI rotation curves. The SPARC sample contains disc galaxies with a broad range of luminosity, surface brightness, rotation velocity, and Hubble type. It therefore forms a representative sample of the nearby Universe. Two main results from the SPARC data are used in the following section. Firstly, \citet{SPARC} found scaling relations between several characteristics of a disc galaxy, e.g between the stellar and HI mass. Secondly, there is a very tight RAR, which allows us to determine the rotation curve of a galaxy from only its observed baryonic matter distribution \citep{rar}. The RAR relates the observed acceleration to the Newtonian acceleration of the baryons alone, so we use it to calculate the rotation curve $V \left( R_{\text{gal}} \right)$. The RAR has an obvious interpretation in MOND, while in a Newtonian context, it must capture the extra gravity due to the dark halo.

Assuming the scaling relations from SPARC \citep{SPARC} and the RAR \citep{rar}, one can use the here developed star formation law to make further predictions. For simplicity, we make the following assumptions:
\begin{enumerate}
	\item The mass-to-light ratio of the stellar population is 0.5 $M_{\odot}/L_{\odot}$ at 3.6 $\mu$m \citep{Schombert_2014}.
	\item The total gas mass in the galaxy is 1.33 times the HI mass ($M_{\text{HI}}$).
	\item Both the stellar and the gas disc are single exponential discs with scale length $R_d$ taken from the SPARC scaling relations (see below).
	\item The scatter of the SPARC scaling relations is ignored here.
\end{enumerate}
With these simplifications (masses being in units of $M_{\odot}$ and radii in kpc), we obtain \citep[eq. 4 of][]{SPARC}:
\begin{equation}
	\log_{10} M_{\text{HI}} ~=~ 0.54 \log_{10} M_S + 4.06\, ,
	\label{MHI_MS}
\end{equation}
with $M_S$ being the stellar mass. Combining eqs. 3 and 6 from \citet{SPARC} with their eq. 3 corrected to have $+7.20$ rather than $-7.20$ yields the second scaling relation we use:
\begin{equation}
	\log_{10} R_d ~=~ 0.62 \log_{10} M_{\text{HI}} - 5.40 \, .
	\label{Rd_MHI}
\end{equation}

To calculate $V \left( R_{\text{gal}} \right)$, we first determine the Newtonian rotation curve of the combined stellar ($V_{\text{disc,stellar}}(R_{\text{gal}})$) and gas disc ($V_{\text{disc,gas}}(R_{\text{gal}})$). We combine them in quadrature to get the total Newtonian rotation curve $V_{\text{N}}(R_{\text{gal}})$:
\begin{equation}
	V_{\text{N}}(R_{\text{gal}}) ~=~ \sqrt{V_{\text{disc,stellar}}^2 \left( R_{\text{gal}} \right) + V_{\text{disc,gas}}^2 \left( R_{\text{gal}} \right)} \, .
\end{equation}
We then use the RAR \citep[eq. 11 of][]{rar} to convert this into $V \left( R_{\text{gal}} \right)$. With the above two scaling relations and $V \left( R_{\text{gal}} \right)$ determined via the RAR, we can calculate $\Sigma_{\text{SFR}}$ in both Newtonian and MONDian disc galaxies.

While the relation between radial gravity and baryonic distribution is the same in both theories, there is a fundamental difference in that the Newtonian discs are not self-gravitating, whereas the MONDian ones are. This leads to differences in their stability and secular evolution, which are also partly driven by the dark matter halo being present in only the Newtonian context \citep[e.g.][]{Banik_2020_M33, Roshan_2021}.

\subsection{Comparison of the SFR in Newtonian and MOND gravity}

\begin{figure}
	\begin{center}
		\includegraphics[width=85mm,height=75mm]{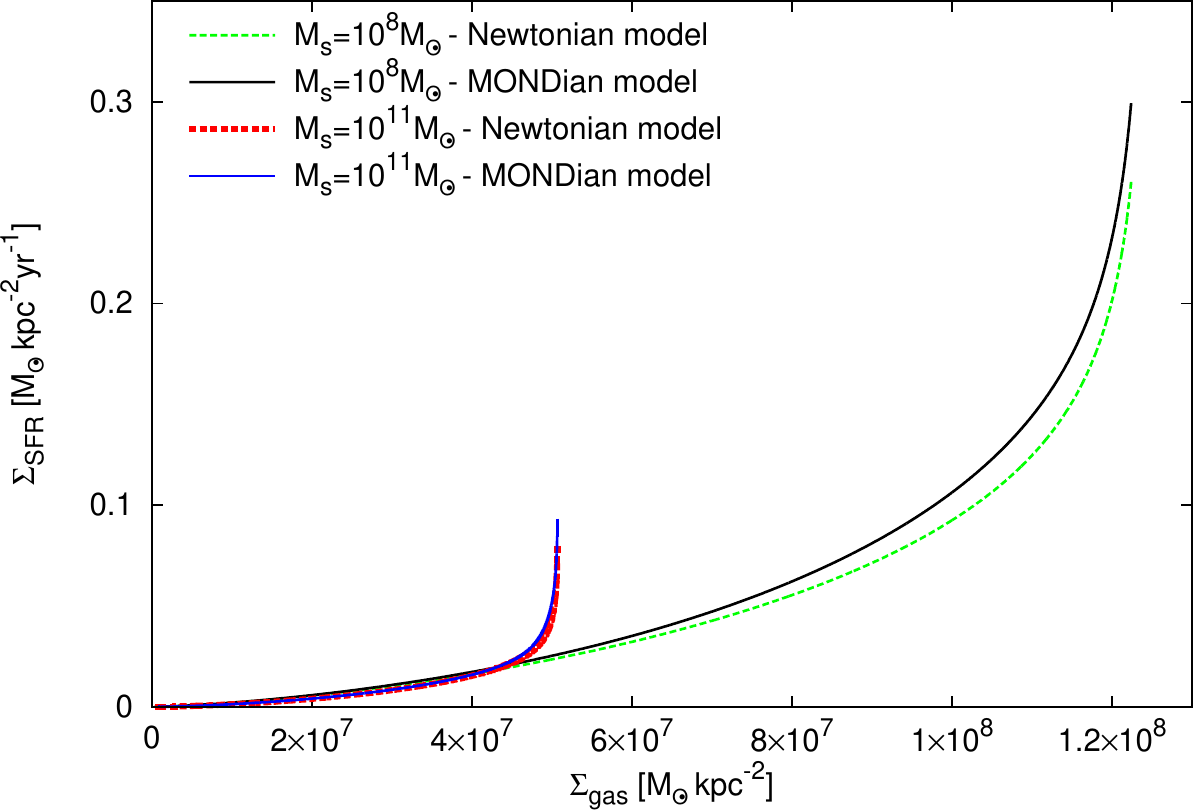}
		\caption{$\Sigma_{\text{SFR}}$ for the Newtonian and MONDian models as a function of $\Sigma_{\text{gas}}$ for two galaxies with $M_S = 10^8$ $M_{\odot}$ and $M_S = 10^{11}$ $M_{\odot}$. The green curve is the Newtonian $\Sigma_{\text{SFR,N}}$ for a galaxy with  $M_S = 10^8$ $M_{\odot}$, using the SPARC scaling relations \citep{SPARC} including the RAR \citep{rar} and ignoring scatter. The black line is the MONDian $\Sigma_{\text{SFR,M}}$. The red line is the Newtonian model for $M_S = 10^{11}$ $M_{\odot}$, while the blue line is the same for the MONDian model. The graph shows that the difference between the models is small, but much more noticeable in the low-mass galaxy. Note that although this galaxy has a higher local $\Sigma_{\text{SFR}}$ than the more massive galaxy, its smaller size (Eq. \ref{Rd_MHI}) means its total SFR is still much lower (see also Fig. \ref{main_seq}).}
		\label{sfrshi}
	\end{center}
\end{figure}

In Fig. \ref{sfrshi}, we show the calculated $\Sigma_{\text{SFR,M}}$ and $\Sigma_{\text{SFR,N}}$ in dependence of $\Sigma_{\text{gas}}$ for two galaxies with galactic stellar mass $M_S = 10^8$ $M_{\odot}$ and $M_S = 10^{11}$ $M_{\odot}$. The low mass galaxy is gas-dominated with higher gas density in the central regions. The Newtonian and MOND results are very similar.

\begin{figure}
	\includegraphics[width=88mm,height=75mm]{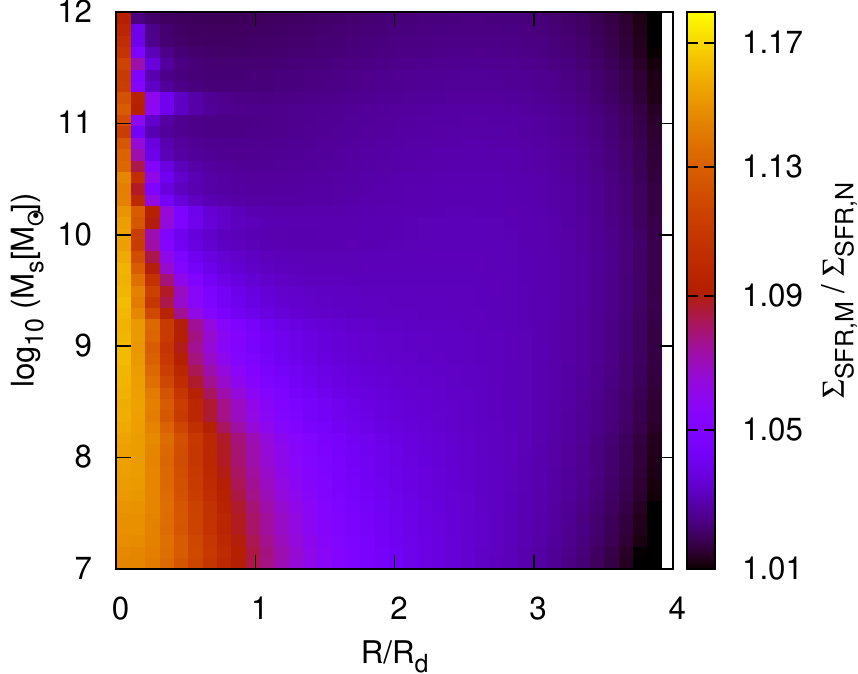}
	\caption{Ratio of the MOND $\Sigma_{\text{SFR}}$ (Eq. \ref{SSFRfinal_gen}) to the Newtonian result (Eq. \ref{Ssfrfinal}), assuming the SPARC scaling relations (Section \ref{SPARC_scaling_relations}) hold without scatter. The $x$-axis shows galactocentric distance in units of the scale length $R_d$, while the $y$-axis shows the logarithm of the stellar mass $M_S$ of the galaxy in Solar masses.}
	\label{ssfr_gal}
\end{figure}

It is possible to show how much star formation is boosted in different regions of different galaxies by comparing MONDian to Newtonian dynamics. The relative deviation of $\Sigma_{\text{SFR,M}}$ from $\Sigma_{\text{SFR,N}}$ is calculated using Eqs. \ref{SSFRfinal_gen} and \ref{Ssfrfinal} adopting different values for $g_{\text{ext,N}}$ and $g_{\text{int,N}}$ (Fig. \ref{ssfr_dev}). As can be seen, $\Sigma_{\text{SFR,M}}$ shows the maximum deviation in the idM regime where $g_{\rm{ext}} \ll g_{\rm{int}} \ll a_0$. In addition, Fig. \ref{ssfr_gal} shows $\Sigma_{\text{SFR,M}}/\Sigma_{\text{SFR,N}}$ as a function of galactocentric distance ($R/R_d$) and $M_S$. The central parts of all galaxies lie in the deep-MOND regime, which shows the maximum 20\% deviation. Low mass galaxies go directly from the deep-MOND limit to the quasi-Newtonian limit, whereas high-mass galaxies go first into the Newtonian limit as they typically have a higher surface density. Therefore, MOND gives less massive galaxies a bigger boost to their star formation rate.

\begin{figure}
	\includegraphics[width=85mm,height=70mm]{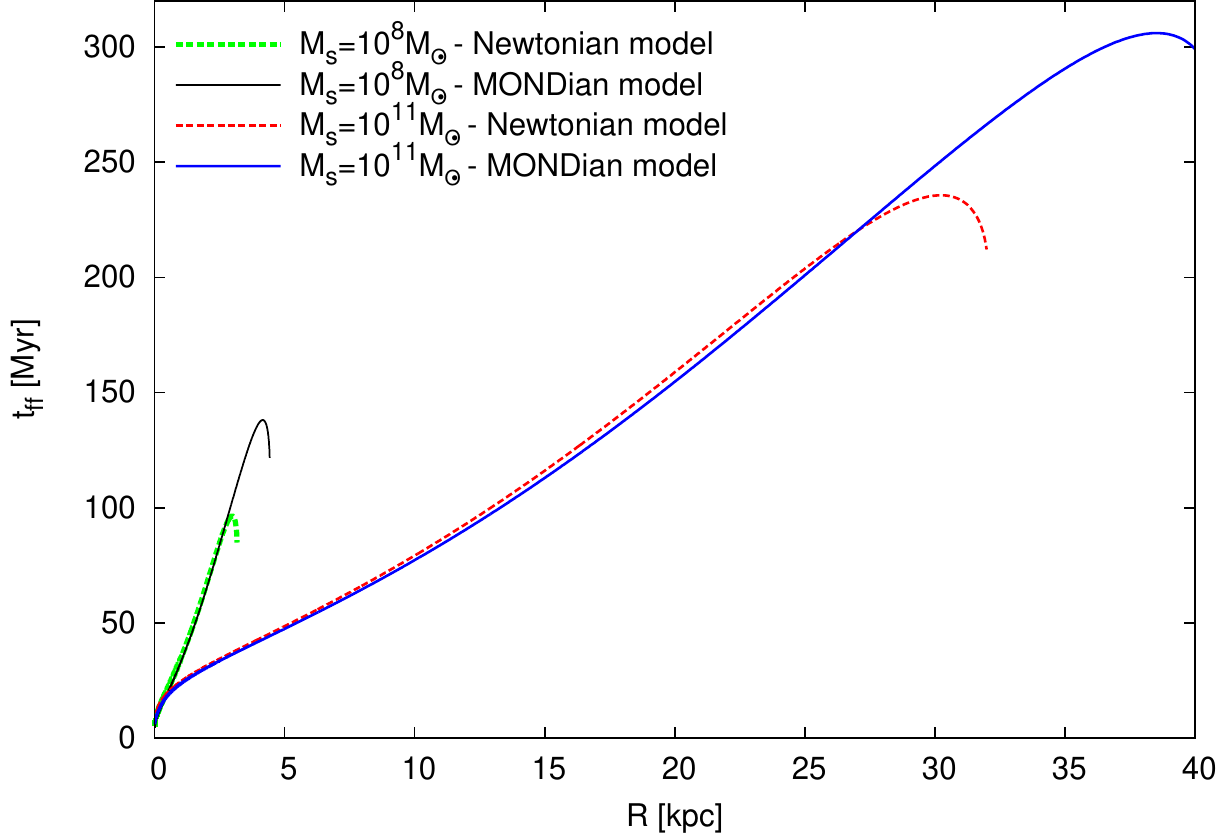}
	\caption{The free-fall time as a function of galactocentric distance in the Newtonian gravity (dashed lines) and MOND (solid lines) for two galaxies with $M_S = 10^8 M_{\odot}$ (green and black lines) and $M_S = 10^{11} M_{\odot}$ (red and blue lines). Notice that star formation extends out to larger radii in MOND.}
	\label{tff_gal}
\end{figure}

Comparing the free-fall time as a function of galactocentric distance in Fig. \ref{tff_gal} for the MONDian and Newtonian models, we see that star formation can extend to larger radii in MOND. This is because the outer regions of a galaxy are in the low acceleration regime, where MOND significantly enhances the disc self-gravity and thus makes it more likely to gravitationally collapse. For a numerical study into star formation in MOND gravity, we refer the reader to \citet{Renaud_2016}.

\begin{figure}
	\begin{center}
	\includegraphics[width=85mm,height=75mm]{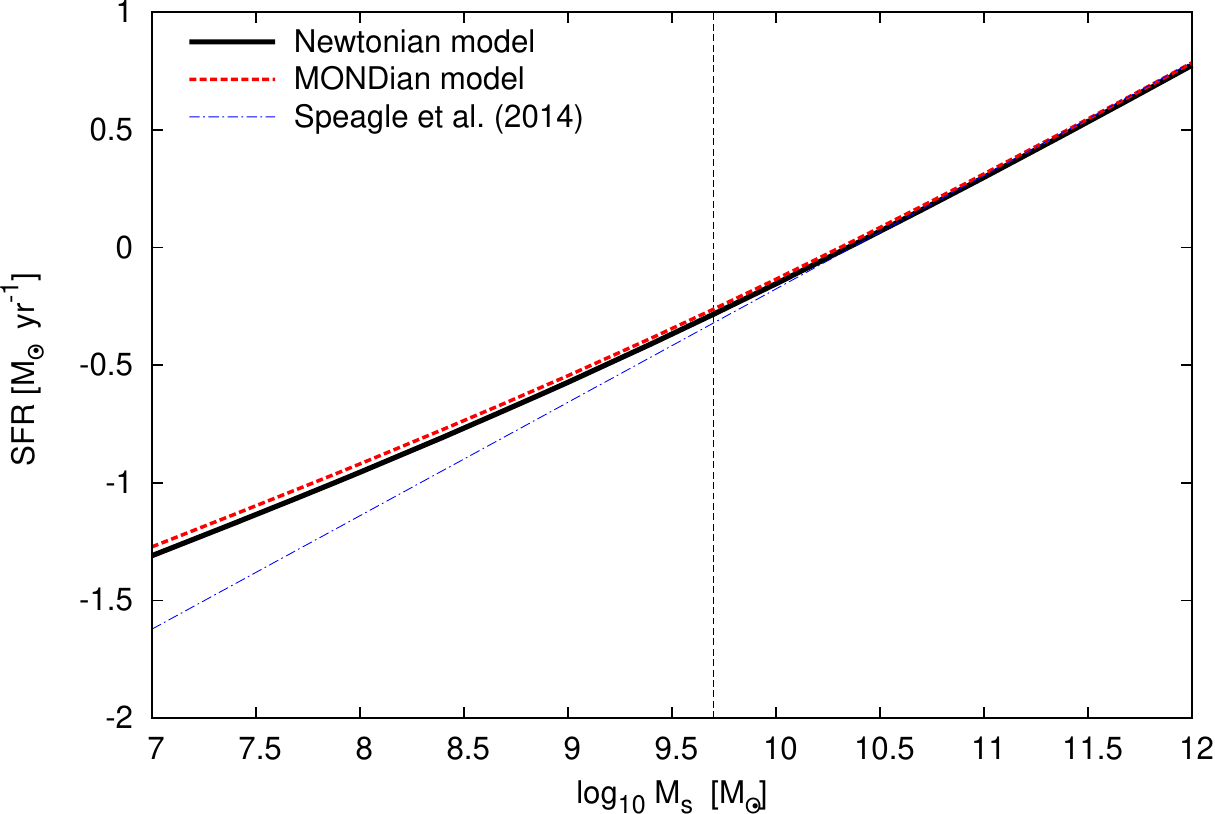}
	\caption{The total star formation rate in comparison to the stellar mass of a galaxy. The dotted blue line is a fit to observations by \citet{Speagle} using an age of the universe of 13.77 Gyr, though note that the regression is based only on galaxies with stellar mass higher than indicated by the vertical dashed line. The black curve shows the numerical integral over the whole galaxy of Eq. \ref{Ssfrfinal} using the SPARC scaling relations (Section \ref{SPARC_scaling_relations}). Therefore, the black line represents the total SFR in Newtonian dynamics. The red line is the same for MOND (Eq. \ref{SSFRfinal_gen}). In both cases, we use a constant star formation efficiency of $\varepsilon_{\text{SFE}} = 1.1\%$ as it best fits the observed main sequence. The derived relations are very similar in both gravity laws, and for the most part closely follow the observed relation. Only for the low-mass regime do they begin to diverge.}
	\label{main_seq}
	\end{center}
\end{figure}

There have also been several observational studies exploring the relation between the stellar mass of a galaxy and its SFR, the so-called main sequence of galaxies \citep[e.g.][]{Speagle}. This can be compared with the results of the here developed theory. Assuming a constant $\varepsilon_{\text{SFE}}$, it is possible to radially integrate the here developed star formation law using Eq. \ref{Ssfrfinal} (Eq. \ref{SSFRfinal_gen}) to get the total SFR in Newtonian (Milgromian) dynamics. For both cases, we use a constant star formation efficiency of $\varepsilon_{\text{SFE}} =1.1\%$ as it best fits the observed main sequence. A comparison with the here calculated main sequence and the one from \citet{Speagle} can be seen in Fig. \ref{main_seq}. Although low-mass galaxies have a higher local $\Sigma_{\text{SFR}}$ than more massive galaxies, the total integrated SFR is still much lower because the galaxies are smaller (Eq. \ref{Rd_MHI}). 

Note that the observed main sequence of galaxies is valid only at $M_S > 10^{9.5} M_{\odot}$. At low $M_S$, our results show a deviation from the extrapolated observed curve derived by \citet{Speagle}. This can be interpreted as evidence for deviation of low-mass galaxies from the observed main sequence defined by more massive galaxies \citep{Kroupa_2020}.  Alternatively, if the SFE of galaxies is a function of stellar mass such that a lower mass galaxy  would have a lower value of the SFE, then the low mass galaxies would reach a lower SFR compared to that calculated here for a constant SFE in all galaxies. Another possible complication is the integrated galactic initial mass function \citep[IGIMF;][]{KroWei} theory according to which the galaxy-wide IMF becomes top-light at small SFRs (see also \citealt{Jerabkova_2018}). The IGIMF-based analysis of dwarf galaxy SFRs by \citet{Pflamm_2009} indicates that the main sequence extends to $M_s \approx 10^7 M_{\odot}$.

\begin{figure*}
	\begin{center}
	\includegraphics[width=80mm,height=67mm]{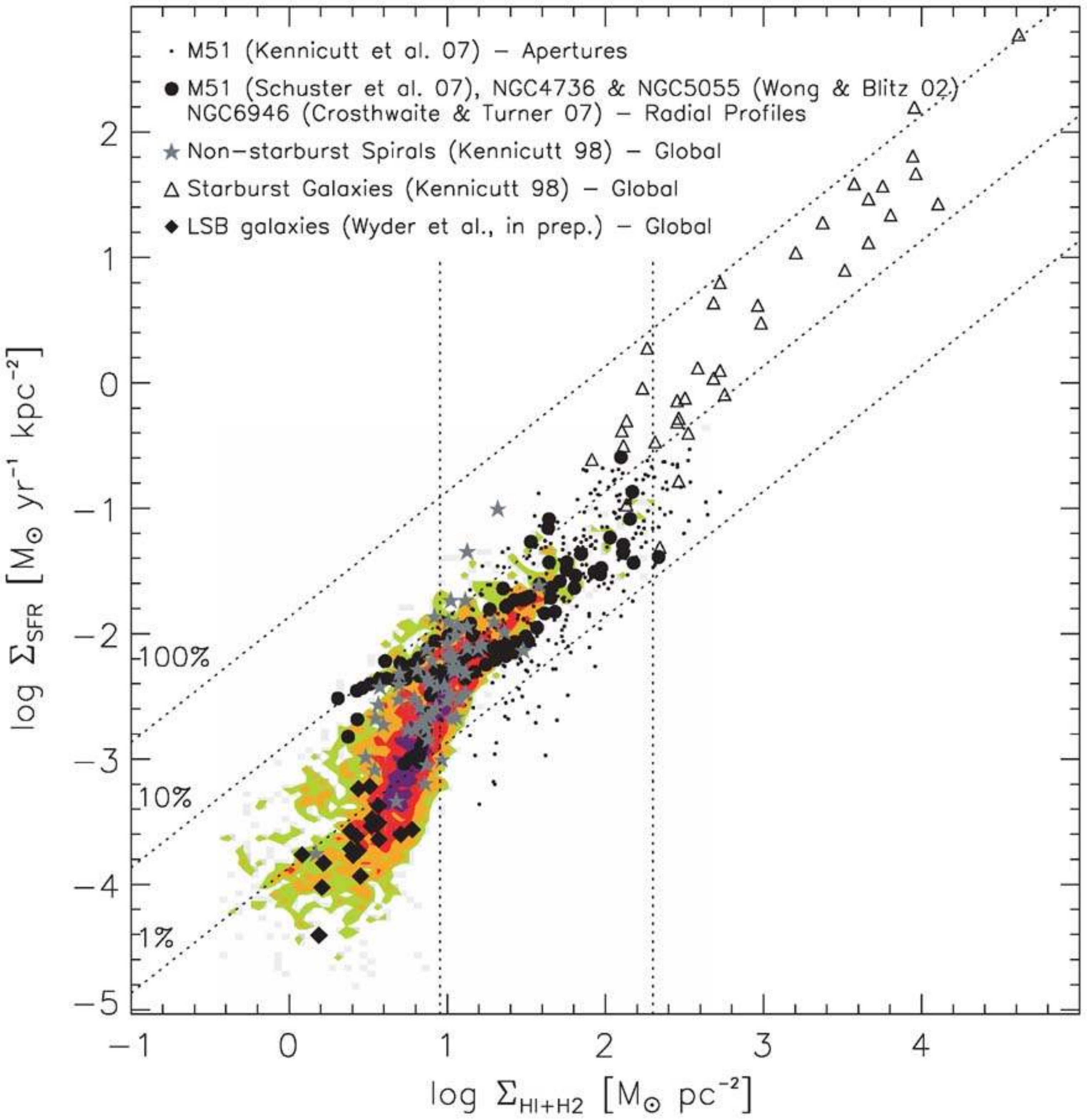}
	\includegraphics[width=80mm,height=70mm]{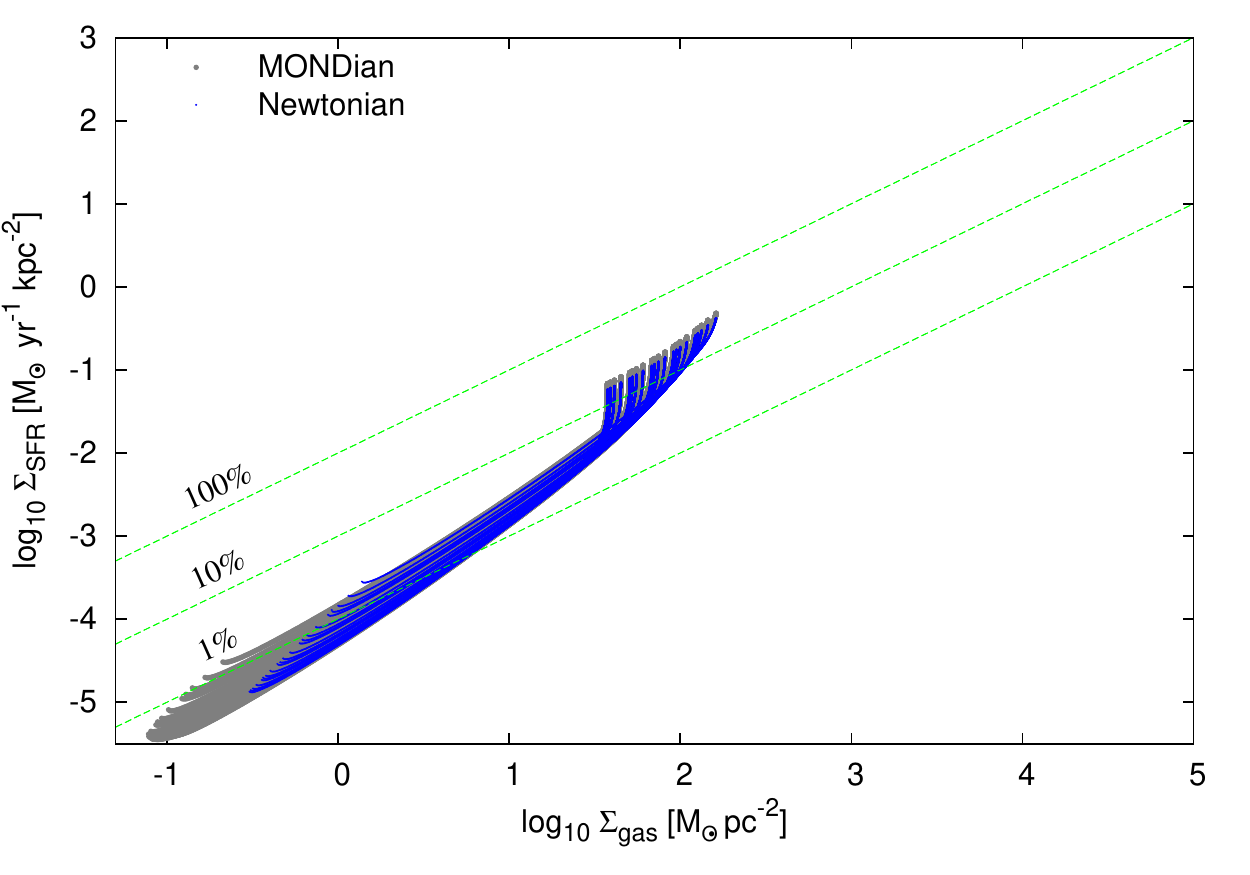}
	\caption{The relation between $\Sigma_{\text{SFR}}$ and $\Sigma_{\text{gas}}$. \emph{Left}: fig. 15 from \citet{Bigiel}, which contains observational data from their work (coloured dots), and further data as mentioned in the legend. \emph{Right}: The results of our model for a variety of galaxies following the SPARC scaling relations, shown using small blue dots for Newtonian gravity and larger grey dots for MOND that extend to lower values. The dots represent local values of $\Sigma_{\text{SFR}}$ in different galaxies assuming  $\varepsilon_{\text{SFE}}=1.1$\%. The diagonal dotted lines in both panels show lines of constant SFE, indicating the level of $\Sigma_{\text{SFR}}$ needed to consume 1\%, 10\%, and 100\% of the gas reservoir in $10^8$ years. Thus, the lines also correspond to constant gas depletion times of, from top to bottom, $10^8$, $10^9$, and $10^{10}$ years. Our results do not follow these lines because the free-fall time is not always the same (see the text).}
	\label{sigma}
	\end{center}
\end{figure*}

\citet{Bigiel} compared various observational data of $\Sigma_{\text{gas}}$ and $\Sigma_{\text{SFR}}$ at sub-kpc resolution taken from the literature for different types of galaxies including starburst galaxies, low surface brightness galaxies, and late-type galaxies, including dwarfs. As can be seen in Fig. \ref{sigma}, the $\Sigma_\text{SFR}-\Sigma_\text{gas}$ relationship varies significantly among and within individual galaxies. Our calculated relationship and those from the literature show a clear, consistent trend in $\Sigma_\text{SFR}-\Sigma_\text{gas}$ space. The right panel of Fig. \ref{sigma} shows that the slope of our calculated $\Sigma_\text{SFR}-\Sigma_\text{gas}$ relation is about 1.4. This is steeper than the $n=1$ assumed in Eq. \ref{Ssfr} because the pre-factor $t_{\rm{ff}}$ depends on both $R_{\rm{gal}}$ and $\Sigma_\text{gas}$ (see Eqs. \ref{Free_fall_time_Newton} and \ref{Rcloud}). The fact that our model predicts $n>1$ can also be extracted from the combination of Eqs. \ref{Ssfr} and \ref{Free_fall_time_Newton}. It is caused by the smaller free-fall time in regions with a higher surface density, and is therefore a very robust prediction of our model for both Newtonian and Milgromian gravity.

It should be noted that there are some data for starburst galaxies with high $\Sigma_{\text{gas}}$ and $\Sigma_{\text{SFR}}$ that cannot be reproduced in our formalism. Such galaxies are out of equilibrium as they are typically interacting with another galaxy, so they are not accounted for by the present model which assumes galaxies to be in dynamical and self-regulated equilibrium. Also, in the context of the IGIMF theory that predicts the galaxy-wide IMF becomes top-heavy in a high SFR environment, these galaxies should have a top-heavy IMF and form a higher proportion of more massive stars. Therefore, the H$\alpha$ emission applied to estimate the SFR cannot be a good tracer for the SFR of these galaxies $-$ the true SFRs are likely smaller \citep[see fig. 7 of][]{Jerabkova_2018}.

\section{Conclusions}
\label{Conclusions}

The rate of star formation and amount of gas are correlated across a wide sample of star-forming galaxies by the KS law $\Sigma_\text{SFR} \propto \Sigma_\text{gas}^n$. Assuming that a fraction $\varepsilon_\text{SFE}$ of gas is converted into stars every free-fall time $t_\text{ff}$, we calculated $\Sigma_\text{SFR}$ by dividing $\Sigma_\text{gas}$ by the local $t_\text{ff}$. We did this by using linear stability analysis of a uniformly rotating thin disc to determine the size of a collapsing perturbation in Newtonian and Milgromian dynamics \citep{Toomre, MoNDdisc}. We then calculated the initial size and mass of a collapsing cloud as a function of $\Sigma_\text{gas}$ and the rotation curve. Additionally, we found an algebraic approximation for the free-fall time in MOND (Eq. \ref{tffMOND}). In this way, we analytically derived the relation $\Sigma_\text{SFR} \propto \Sigma_\text{gas}^n$ with $n = 1.4$ in both Newtonian and Milgromian dynamics. The general dependence on the variables $\Sigma_\text{gas}$ and $\kappa$ does not differ between the Newtonian and MOND approaches. They differ only in the constant pre-factor, resulting in increased local $\Sigma_\text{SFR}$ by up to 20\% if MOND is valid (Fig. \ref{ssfr_gal}).

We determined the MONDian KS law using well-established physical principles, showing that if star formation is possible, then the influence of the gas temperature is at most a factor of $\sqrt{2}$ (Eq. \ref{finalwithW}). This also applies in Newtonian dynamics. The empirically determined correction factors to the general KS law found by \citet{Prantzos} and \citet{Boissier} are explained here through the epicyclic frequency $\kappa$.

We showed that our calculated $\Sigma_\text{SFR}-\Sigma_\text{gas}$ relationship is in good agreement with the observed present-day main sequence of galaxies, which is consistent with the here derived star formation law $\Sigma_{\text{SFR}} \propto \Sigma_{\text{gas}}^{1.4}$ for $\varepsilon_\text{SFE} = 1.1 \%$. It is worth mentioning that the $\Sigma_\text{SFR}-\Sigma_\text{gas}$ relation might be shallower once IGIMF effects are taken into account, since traditional measures of the SFR would be underestimates (at small SFRs) or overestimates (at high SFRs, \citealt{Jerabkova_2018}). The resulting higher SFR in dwarf galaxies \citep*{Pflamm_2009, Pflamm-etal_2009} could be accounted for in our model with a higher star formation efficiency, which is likely in the case of having less feedback. 







\section*{Acknowledgements}

AHK is supported by an Alexander von Humboldt Foundation postdoctoral research fellowship. IB is supported by an Alexander von Humboldt Foundation postdoctoral research fellowship and the ``Pathways to Research'' programme from the University of Bonn. PK acknowledges support from the Grant Agency of the Czech Republic under grant number 20-21855S.

\section*{Data availability}
The data underlying this article are available in the article.

\bibliographystyle{mnras}
\bibliography{KSM_bbl}

\begin{appendix}

\section{The free-fall time in MOND}
\label{tff_derive}

In Newtonian dynamics, there are two ways to calculate the free-fall time. The first is to start from the equation of motion in free fall:
\begin{equation}
	\frac{d^2r}{dt^2} ~=~ -\frac{GM}{r(t)^2} \, ,
\end{equation}
and solve the differential equation in time $t$. Another possibility without referring to differential equations would be Kepler's Third Law. Assuming that one has two objects orbiting each other with a semi-minor axis of 0 and a semi-major axis of $R/2$ with $R$ being the initial distance of the two masses, we would get the same result since the free-fall time would be half the orbital period.

In MOND, there are two things to keep in mind: the first approach, using the equation of motion, can only be done for the extreme cases, as the differential equation is not algebraically solvable in the general case (Eq. \ref{alg_approx}):
\begin{equation}
	\frac{d^2r}{dt^2}=-\frac{GM}{r(t)^2}\nu\left( \frac{g_{\text{N}}(t)}{a_0} \right) \left(1+\tanh\left(0.825 \frac{g_{\text{ext,N}}r(t)^2}{GM} \right)^{3.7} \frac{K(t)}{3}\right) \, ,
\end{equation}
here with added emphasis on the time-dependent parts. Secondly, the approach using Kepler's Third Law leads to a different result compared to the first approach. In the case of the idM limit, the numerical difference between the two approaches reaches its maximum of $2/\sqrt{\mathrm{\pi}}$. The reason for this is that MOND gravity from a point mass is in general not spherically symmetric \citep{Banik_2015}, so the approach using Kepler's Third Law yields an offset to the correct value. This offset disappears in the EFE-dominated or Newtonian cases.

The important point is that the approach using Kepler's Third Law can be used for any intermediate case, whereas the approach using a differential equation can only be used for the extreme cases. Using Kepler's Third Law gives the correct dependencies and is only off by a numerical factor of order unity. Therefore, one can use this approach to write the free-fall time in MOND as:
\begin{equation}
	t_{\text{ff}}=\frac{\mathrm{\pi}}{2}\frac{R^{3/2}}{\sqrt{2 G M \nu \left( g_{\text{N}}/a_0 \right) \left(1+\tanh\left(0.825 \frac{g_{\text{ext,N}}}{g_{\text{int,N}}} \right)^{3.7} \frac{K}{3}\right)}} N \, ,
	\label{tffMOND_app}
\end{equation}
with $N$ being a numerical factor between 1 and $2/\sqrt{\mathrm{\pi}}$. We can think of $N$ as an interpolating function with the following characteristics: $N=1$ in the Newtonian and EFE-dominated cases, and $N=2/\sqrt{\mathrm{\pi}}$ in the idM limit to reproduce eq. 24 of \citet{MoNDfft}. As a compromise between accuracy and simplicity, we found the following function gives a maximum error of 3\%:
\begin{equation}
	N ~=~ 1 + \frac{\left( 1-\frac{1}{\nu \left( g_{\text{N}}/a_0 \right)} \right) \left( \frac{2}{\sqrt{\mathrm{\pi}}}-1 \right)}{\frac{g_{\text{ext,N}}}{g_{\text{int,N}}}+1}  \, .
	\label{N_app}
\end{equation} 
It is possible to find even more accurate interpolating functions, albeit more complicated ones. For example, the maximum error can be reduced to only 1.2\% using:
\begin{equation}
	N ~=~ 1 + \frac{\left( 1-\frac{1}{\nu \left( g_{\text{N}}/a_0 \right)} \right) \left( \frac{2}{\sqrt{\mathrm{\pi}}} - 1 \right)}{\left(\frac{g_{\text{ext,N}}}{g_{\text{int,N}}}\right)^{0.6} \ln\left(\frac{g_{\text{ext,N}}}{g_{\text{int,N}}} +1 \right) + 1} \, .
	\label{N_app_comp}
\end{equation}

\end{appendix}

\bsp
\label{lastpage}
\end{document}